\documentclass[12pt, draftclsnofoot, onecolumn]{IEEEtran}
\usepackage{amsfonts}
\usepackage{amsmath}
\usepackage{amssymb}
\usepackage{bm}
\usepackage{xcolor,graphicx,float}
\usepackage{color, soul}
\usepackage{stfloats}
\usepackage[numbers,sort&compress]{natbib}
\usepackage[amsmath,thmmarks]{ntheorem}
\usepackage{theorem}
\usepackage{algorithm}
\usepackage{algorithmic}
\usepackage{graphicx}
\usepackage{subfigure}
\usepackage{pict2e}
\newtheorem{theorem}{Theorem}

\theoremheaderfont{\sc}\theorembodyfont{\upshape}
\theoremstyle{nonumberplain}
\theoremseparator{}
\theoremsymbol{\rule{1ex}{1ex}}

\begin{document}
\title{Gridless Variational Bayesian Channel Estimation for Antenna Array Systems with Low Resolution ADCs}

\author{Jiang~Zhu, Chao-kai Wen, Jun Tong, Chongbin Xu and Shi Jin\thanks{J. Zhu is with Ocean College, Zhejiang University, Zhoushan 316021, China (e-mail: jiangzhu16@zju.edu.cn). C.-K. Wen is with the Institute of Communications Engineering,
National Sun Yat-sen University, Kaohsiung 80424, Taiwan (e-mail:
chaokai.wen@mail.nsysu.edu.tw). J. Tong is with School of Electrical, Computer and Telecommunications Engineering, University of Wollongong, Wollongong, NSW 2522, Australia (e-mail: jtong@uow.edu.au). C. Xu is with Key Laboratory for Information Science of Electromagnetic Waves (MoE), the Department of Communication Science and Engineering, Fudan University,
Shanghai 200433, China (e-mail: chbinxu@fudan.edu.cn). S. Jin is with the National Mobile Communications Research Laboratory,
Southeast University, Nanjing 210096, China (e-mail: jinshi@seu.edu.cn).}}
%\thanks{Jiang Zhu and Qi Zhang are with the key laboratory
%of ocean observation-imaging testbed of Zhejiang Province, Ocean College,
%Zhejiang University, No.1 Zheda Road, Zhoushan, 316021, China. Xiangming Meng is with Huawei Technologies, Co. Ltd., Shanghai, 201206, China. }
%\thanks{Jiang Zhu and Qi Zhang are with Ocean College, Zhejiang University. Xiangming Meng is with Huawei company. Jun Fang is with the University of Electronic Science and Technology of China. }
%\date{}
%{\center\today}
\maketitle
\begin{abstract}
Employing low-resolution analog-to-digital converters (ADCs) coupled with large antenna arrays at the receivers has drawn considerable interests in the millimeter wave (mm-wave) system. Since mm-wave channels are sparse in angular dimensions, exploiting the structure could reduce the number of measurements while achieve acceptable performance at the same time. Motivated by the variational Bayesian line spectral estimation (VALSE) algorithm which treats the angles as random parameters, in contrast with previous works which confine the estimate to the set of grid angle points and induce grid mismatch, this paper proposes the grid-less quantized variational Bayesian channel estimation (GL-QVBCE) algorithm for antenna array systems with low resolution ADCs. Compared to the traditional least squares (LS) approach, numerical results show that GL-QVBCE performs significantly better and asymptotically approaches the Cram\`{e}r Rao bound (CRB).
\end{abstract}
%\begin{keywords}
{\bf keywords}: Variational Bayesian inference, expectation propagation, quantization, grid-less, MMSE, LMMSE, CRB
%\end{keywords}
\section{Introduction}
As the spectrum available is limited in sub-6 GHz bands, millimeter wave (mm-wave) communications have drawn a great deal of attention as a potential technology for future cellular system \cite{Rappaport}. Since mm-wave communication suffers from high attenuation, large antennas are often employed at the receiver side to focus their power. As the bandwidth scales up, the power consumption and implementation complexity of conventional analog-to-digital converters (ADCs) (8-14bits) increase significantly \cite{Singh1}. Consequently, low resolution ADCs (1-3bits) are proposed to be employed at the receiver side \cite{Rangan1}.

Since low-resolution ADCs incur severe distortion on the original received signal, traditional channel estimation algorithms may suffer significant performance degradation. Consequently, many studies have proposed novel signal processing algorithms to estimate the channel from heavily quantized samples \cite{Mo, Choi}. To reduce the training cost, a joint channel-and-data estimation method is proposed \cite{Bayesjoint}, although with high computation complexity.

The mm-wave channel is often characterized by a single dominant path and several other weaker paths \cite{RappaportTcom, Samimi, Rodriguez}. In \cite{gridCE}, a grid-based compressive sensing approach which confines the direction of arrivals (DOAs) to the set of grid angle points is proposed. Since the DOAs are continuous parameters, such an approach incurs basis mismatch and becomes more significant as the bit-depth decreases. To address the basis mismatch issue, a grid-less atomic norm based channel estimation approach is proposed under mixed one-bit antenna systems \cite{Wen}.  As shown in \cite{Wen}, exploiting this structure improves the channel estimation performance significantly over conventional channel estimators such as maximum likelihood (ML) and linear minimum mean squared error (LMMSE) estimators. Since the atomic norm based approach involves solving a semidefinite programming (SDP), the computation complexity is high for large antenna systems.

The angular models of the propagation channels for mm-wave communication share the similar structure as the line spectral estimation (LSE) or DOA problem. Consequently, advanced array processing methods may help address the channel estimation problem in wireless communications. In \cite{Badiu}, a variational Bayesian LSE (VALSE) is proposed to perform the LSE. The VALSE treats the frequencies as random parameters, and it automatically estimates the model order (number of spectral), noise variance and the nuisance parameters of the prior distribution. In addition, VALSE also outputs the posterior probability density fucntion (PDF) of the frequencies and provides the uncertain degree of frequency estimates, in contrast with the prior work providing only point estimates. Later, multisnapshot VALSE (MVALSE) is developed to deal with the multiple measurement vector (MMV) \cite{JiangVALSE}. In \cite{JiangVALSEQ}, VALSE is extended to solve the LSE from quantized samples. In this work, we derive the VALSE solution to the channel estimation problem.

%Performing the channel estimation via exploiting the angular structures from quantized samples is more challenging as it involves two nonlinear transforms, one comes from the angular model, another from the heavy quantization. In \cite{MengZhu}, A grid-based generalized SBL (Gr-SBL) algorithm is proposed to solve the DOA from one-bit samples. Since the DOAs are confined to lie on the grids, the problem can be abstracted as the generalized linear model (GLM).

From the algorithm point of view, many standard Bayesian algorithms \cite{AMP, Wipf, guo, TurboSR, OAMP, GGAMPSBL, VAMP} such as approximate message passing (AMP) \cite{AMP}, sparse Bayesian learning (SBL) \cite{Wipf}, orthogonal approximate message passing (OAMP) \cite{OAMP}, vector approximate message passing (VAMP) \cite{VAMP} have been proposed to deal with the standard linear model (SLM). For the generalized linear model (GLM), Bayesian algorithms \cite{GAMP, GVAMP, JinGTR, JinGEC, meng1, MengZhu, zhucomemnt, AMPGrSBL} such as generalized AMP (GAMP) \cite{GAMP}, generalized expectation consistent signal recovery (GEC-SR) algorithm \cite{JinGEC}, and generalized VAMP \cite{GVAMP} are developed. Later, a unified Bayesian inference framework is proposed, which demonstrates that the GLM can be solved via exchanging extrinsic information between the SLM and the minimum mean square error (MMSE) module \cite{meng1}. Then, generalized AMP (Gr-AMP), generalized VAMP (Gr-VAMP), and generalized SBL (Gr-SBL) are developed. In addition, by discreting the DOAs into the grids, Gr-SBL is applied to solve the DOA from one-bit samples \cite{MengZhu}. For the inference in deep networks, multi-layer VAMP (ML-VAMP) derived from expectation propagation \cite{Minka} is proposed, and the mean-squared error performance of ML-VAMP is exactly predicted in a certain large system limit \cite{multilayer}. As shown later, the channel estimation in this paper can be viewed as a problem of estimating the line spectral which undergoes a linear transform followed by a componentwise nonlinear transform. As a result, we could utilize EP to design algorithms by implementing the respective modules and scheduling the messages between these modules \footnote{For more details about EP and its relation to pmessage passing algorithms, please refer to \cite{AMPEP, Wu, Haochuan, VAMP, meng1, AMPGrSBL}}.

This paper proposes a grid-less quantized variational Bayesian channel estimation (GL-QVBCE) approach. The GL-QVBCE is designed from the module point of view. To evaluate the performance of the GL-QVBCE, we also use the additive quantization noise model (AQNM) and design GL-VBCE algorithm. Besides, the Cram\`{e}r-Rao bound (CRB) is derived under low resolution ADCs to be acted as the benchmarks of the proposed algorithm. Since the DOAs may be fixed across the pilots, while the complex amplitudes may be varied, sequential GL-QVBCE (Seq-GL-QVBCE) algorithm is proposed to track the channel.  Finally, substantial numerical experiments are conducted to illustrate the effectiveness of the proposed approach, and investigate the factors on channel estimation performance, including signal to noise ratio (SNR), bit-depth, number of effective antennas, number of pilots, etc..

\subsection{Notation}
For a matrix $\mathbf A$, let $[{\mathbf A}]_{i,j}$ or $A_{ij}$ denote the $(i,j)$th element of ${\mathbf A}$, and ${\rm diag}(\mathbf A)$ returns a vector with elements being the diagonal elements of $\mathbf A$. Let $\Re\{\cdot\}$ and $\Im\{\cdot\}$ denote the real and imaginary part operator, respectively. Let $\otimes$ denote the Kronecker product operator. For a vector ${\mathbf a}$, let ${\rm diag}({\mathbf a})$ return a diagonal matrix with diagonal elements being $\mathbf a$. The symbol $t$ refers to the time or iteration index. Let ${\mathcal S}\subset \{1,\cdots,N\}$ be a subset of indices. For a square matrix ${\mathbf A}\in {\mathbb C}^{N\times N}$, let $\mathbf A_{{\mathcal S},{\mathcal S}}$ denote the submatrix by choosing both the rows and columns of $\mathbf A$ indexed by $\mathcal S$. Let ${(\cdot)}^{*}_{\mathcal S}$, ${(\cdot)}^{\rm T}_{\mathcal S}$ and ${(\cdot)}^{\rm H}_{\mathcal S}$ be the conjugate, transpose and Hermitian transpose operator of ${(\cdot)}_{\mathcal S}$, respectively. For a random vector $\mathbf x$ with probability density function (PDF) $p({\mathbf x})$, let ${\rm Proj}[p({\mathbf x})]$ denote the projection of $p({\mathbf x})$ onto Gaussian PDF with diagonal covariance matrix, where the means and variances are matched with that of $p({\mathbf x})$.
\section{System Model}
Consider a pilot signal $x_t\in {\mathbb C}$ that impinges on a linear array with $M$ antennas. The received signal vector $\bar{\mathbf y}_t$ at the $t$th time-instant is expressed as \cite{Wen}
\begin{align}\label{unqmodel}
\bar{\mathbf y}_t=\left(\sum\limits_{l=1}^L\beta_l{\mathbf a}(\phi_l)\right)x_t+{\mathbf w}_t,
\end{align}
where $L$ is the number of rays, $\beta_l\in {\mathbb C}$ is the $l$th ray coefficient, ${\mathbf a}(\phi_l)\in{\mathbb C}^M$ is the array manifold vector with $\phi_l\in [-\pi,\pi)$ as the DOAs of the $l$th ray, ${\mathbf w}_t\in {\mathbb C}^M$ is the additive white Gaussian noise with mean zero and covariance matrix $\sigma^2{\mathbf I}_M$. The array manifold vector ${\mathbf a}(\phi_l)$ for the linear array is
\begin{align}
{\mathbf a}(\phi_l)=[{\rm e}^{{\rm j}m_1\pi\sin(\phi_l)},{\rm e}^{{\rm j}m_2\pi\sin(\phi_l)},\cdots,{\rm e}^{{\rm j}m_M\pi\sin(\phi_l)}]^{\rm T},
\end{align}
where ${\mathcal M}=\{m_1,\cdots,m_M\}\subseteq \{0,1,\cdots,N-1\}$. Note that this model can be applied to general linear arrays.

Denote ${\mathbf A}({\boldsymbol \phi})=[{\mathbf a}(\phi_1),{\mathbf a}(\phi_2),\cdots,{\mathbf a}(\phi_L)]\in {\mathbb C}^{M\times L}$ and ${\boldsymbol \beta}=[\beta_1,\beta_2,\cdots,\beta_L]^{\rm T}$. The channel vector ${\mathbf h}\in {\mathbb C}^M$ can be expressed as
\begin{align}
{\mathbf h}=\sum\limits_{l=1}^L\beta_l{\mathbf a}(\phi_l)={\mathbf A}({\boldsymbol \phi}){\boldsymbol \beta}.
\end{align}
Suppose that ${\mathbf y}_t$ are quantized into a finite number of bits \footnote{As will be shown later, the algorithm can be easily incorporated and extended to work with the mixed ADC system.}, i.e.,
\begin{align}\label{quantmodel}
{\mathbf y}_t={\mathcal Q}(\Re\{\bar{\mathbf y}_t\})+{\rm j}{\mathcal Q}(\Im\{\bar{\mathbf y}_t\}),\quad t=1,\cdots,T.
\end{align}

The goal of this paper is to estimate the channel $\mathbf h$ from the quantized observations $\{{\mathbf y}_t\}_{t=1}^T$. To estimate $\mathbf h$ accurately, the structure of $\mathbf h$ is exploited, i.e., the number of paths $L$, the DOAs $\{\phi_l\}_{l=1}^L$, and their complex gains $\{g_l\}_{l=1}^L$ are estimated.

Combining (\ref{unqmodel}) and (\ref{quantmodel}), and by defining ${\mathbf Y}=[{\mathbf y}_1,\cdots,{\mathbf y}_T]\in {\mathbb C}^{M\times T}$ and the pilot vector ${\mathbf x}=[x_1,\cdots,x_T]^{\rm T}\in {\mathbb C}^T$, one has
\begin{align}
{\mathbf Y}={\mathcal Q}(\Re\{{\mathbf h}{\mathbf x}^{\rm T}+{\mathbf W}\})+{\rm j}{\mathcal Q}(\Im\{{\mathbf h}{\mathbf x}^{\rm T}+{\mathbf W}\}).
\end{align}
Through vectorization and utilize ${\rm vec}({\mathbf A}{\mathbf B}{\mathbf C})=\left({\mathbf C}^{\rm T}\otimes {\mathbf A}\right){\rm vec}({\mathbf B})$, one obtains
\begin{align}\label{singleuser}
{\mathbf y}\triangleq {\rm vec}({\mathbf Y})={\mathcal Q}(\Re\{{\boldsymbol \Phi}{\mathbf h}+{\mathbf w}\})+{\rm j}{\mathcal Q}(\Im\{{\boldsymbol \Phi}{\mathbf h}+{\mathbf w}\}),
\end{align}
where
\begin{align}
{\boldsymbol \Phi}={\mathbf x}\otimes {\mathbf I}_M\in{\mathbb C}^{TM\times M}.
\end{align}
In the following, efficient gridless quantized variational Bayesian channel estimation (GL-QVBCE) algorithm is designed to estimate the channel vector $\mathbf h$. Besides, gridless variational Bayesian channel estimation (GL-VBCE) algorithm can also be designed for the unquantized model.

It is worth noting that the GL-VBCE can be applied to the multiuser scenario. Consider an $K$ user scenario where $K=T$, let ${\mathbf h}_k\in{\mathbb C}^M$ denote the channel between the $k$th user and base station (BS) with the same angular structure as (\ref{unqmodel}), let ${\mathbf x}_k\in {\mathbb C}^T$ denote the pilot of the $k$th user. Since $K=T$, one can construct orthogonal pilots, i.e., ${\mathbf x}_{k_1}^{\rm H}{\mathbf x}_{k_2}=0$, $\forall k_1\neq k_2$. By defining ${\bar Y}=[\bar{\mathbf y}_1,\cdots,\bar{\mathbf y}_T]\in {\mathbb C}^{M\times T}$, we have
\begin{align}
\bar{\mathbf Y}=\sum\limits_{k=1}^K {\mathbf h}_k{\mathbf x}_k^{\rm T}+{\mathbf W}.
\end{align}
For the $k$th user, we extract the $k$th user information by
\begin{align}\label{extuser}
\bar{\mathbf Y}{\mathbf x}_k^{*}&={\mathbf h}_k{\mathbf x}_k^{\rm T}{\mathbf x}_k^{*}+\sum\limits_{k^{'}=1,k\neq k^{'}}^K {\mathbf h}_{k^{'}}{\mathbf x}_{k^{'}}^{\rm T}{\mathbf x}_k^{*}+{\mathbf W}{\mathbf x}_k^{*}\notag\\
&={\mathbf h}_k{\mathbf x}_k^{\rm T}{\mathbf x}_k^{*}+{\mathbf W}{\mathbf x}_k^{*},
\end{align}
which has the form similar to (\ref{unqmodel}). For the quantized setting, if the additive quantization noise model (AQNM) is adopted (see (\ref{aqnmmodel})), then we obtain a single user channel estimation problem similar to equation (9).  While if we directly consider the nonlinear quantization effects, the proposed approach can still be applied. As shown in subsection \ref{moddesign}, the nonlinear model is iteratively approximated as a linear model with noise being heteroscedastic (different components having different variance), we still extract the $k$th user information similarly as (\ref{extuser}) does and perform the channel estimation. The difference is that the extrinsic message between the modules needs to be redesigned. To avoid the obfuscation of the key features of our algorithms by intricate notations, we focus on the single user model (\ref{singleuser}) instead.
\section{GL-QVCBE pAlgorithm}\label{Algorithm}
This section develops the algorithm. First, the probabilistic formulation is introduced. Then, the gridless quantized variational Bayesian channel estimation (GL-QVBCE) algorithm consisting of several different modules are designed.
\subsection{Probabilistic Formulation}
To estimate the channel vector $\mathbf h$ via exploiting the structures, we borrow the probabilistic formulation from \cite{Badiu}. For completeness, we introduce the probabilistic formulation in the following text.

First, we reparameterize the model by defining $\theta_i=\pi\sin \phi_i$ and $\theta_i\in [-\pi,\pi)$. We term $\theta_i$ as the $i$th frequency. Similarly, the number of paths $L$ is termed the number of frequencies. Since  the number of frequencies $L$ is  usually unknown, the number of frequencies is assumed to be $N$ \cite{Badiu}, i.e.,
\begin{align}\label{signal-model}
{\mathbf h}=\sum\limits_{l=1}^N {\beta}_l {\mathbf a}({\theta}_l)\triangleq {\mathbf A}({\boldsymbol \theta}){\boldsymbol \beta},
\end{align}
where ${\mathbf A}({\boldsymbol \theta})=[{\mathbf a}({\theta}_1),\cdots,{\mathbf a}({\theta}_N)]$ and $N$ satisfies $N> L$. Since the number of frequencies is $L$, the binary hidden variables ${\mathbf s}=[s_1,...,s_N]^{\rm T}$ are introduced, where $s_l=1$ means that the $l$th frequency is active, otherwise deactive ($\beta_l=0$). The probability mass function of $s_l$ is
\begin{align}\label{sprob}
p(s_l) = \rho^{s_l}(1-\rho)^{(1-s_l)},\quad s_l\in\{0,1\}.
\end{align}
Given that $s_l=1$, we assume that the channel path coefficient p${\beta}_l\sim {\mathcal {CN}}({\beta}_l;0,\tau)$. Thus $(s_l,{\beta}_l)$ follows a Bernoulli-Gaussian distribution, that is
\begin{align}
p({\beta}_l|s_l;\tau) = (1 - s_l){\delta}({\beta}_l) + s_l{\mathcal {CN}}({\beta}_l;0,\tau).\label{pdfw}
\end{align}

From (\ref{sprob}) and (\ref{pdfw}), it can be seen that the parameter $\rho$ denotes the probability of the $l$th component being active and $\tau$ is a variance parameter. The variable ${\boldsymbol \theta} = [\theta_1,...,\theta_N]^{\rm T}$ has the prior PDF $p({\boldsymbol \theta}) = \begin{matrix} \prod_{l=1}^N p(\theta_l) \end{matrix}$. Generally, $p(\theta_l)$ is encoded through the von Mises distribution \cite[p.~36]{Direc}
\begin{small}
\begin{align}\label{prior_theta}
p(\theta_l) = {\mathcal {VM}}(\theta_l;\mu_{0,l},\kappa_{0,l})= \frac{1}{2\pi{I_0}(\kappa_{0,l})}{\rm e}^{\kappa_{0,l}{\cos(\theta-\mu_{0,l})}},
\end{align}
\end{small}
where $\mu_{0,l}$ and $\kappa_{0,l}$ are the mean direction and concentration parameters of the prior of the $l$th frequency $\theta_l$, $I_p(\cdot)$ is the modified Bessel function of the first kind and the order $p$ \cite[p.~348]{Direc}. Without any knowledge of the frequency $\theta_l$, the uninformative prior distribution $p(\theta_l) = {1}/({2\pi})$ is used \cite{Badiu}. For more details please refer to \cite{Badiu, JiangVALSE}.

By defining
\begin{align}\label{defz}
{\mathbf z}={\boldsymbol \Phi}{\mathbf h},
\end{align}
the PDF $p({\mathbf y}|{\mathbf z})$ of $\mathbf y$ conditioned on $\mathbf z$ can be easily calculated through (\ref{quantmodel}). Let
\begin{align}
&{\boldsymbol \Omega}=(\theta_1,\dots,\theta_N,({\mathbf w},{\mathbf s})),\\
&{\boldsymbol \eta} = \{\rho,~\tau\}
\end{align}
be the set of all random variables and the model parameters, respectively. According to the Bayes rule, the joint PDF $p({\mathbf y },{\mathbf z},{\mathbf h},{\boldsymbol \Omega};{\boldsymbol \eta})$ is
\begin{align}\label{jointpdf}
p({\mathbf y },{\mathbf z},{\mathbf h},{\boldsymbol \Omega};{\boldsymbol \eta})&=p({\mathbf y}|{\mathbf z})\delta_{\rm B}({\mathbf z}-{\boldsymbol \Phi}{\mathbf h})\delta_{\rm A}({\mathbf h}-{\mathbf A}({\boldsymbol \theta}){\boldsymbol \beta})\notag\\
&\times\prod\limits_{i=1}^N p(\theta_i)p(\beta_i|s_i)p(s_i).
\end{align}
Given the above joint PDF (\ref{jointpdf}), the type II maximum likelihood (ML) estimation of the model parameters $\hat{\boldsymbol \eta}_{\rm ML}$ is
\begin{align}\label{MLbeta}
\hat{\boldsymbol\eta}_{\rm ML}=\underset{\boldsymbol \eta}{\operatorname{argmax}}~ p({\mathbf y };{\boldsymbol \eta})=\int p({\mathbf y },{\mathbf z},{\mathbf h},{\boldsymbol \Omega};{\boldsymbol \eta}){\rm d}{\mathbf z}{\rm d}{\mathbf h}{\rm d}{{\boldsymbol \Omega}}.
\end{align}
Then the minimum mean square error (MMSE) estimate of the parameters $({\mathbf z},{\mathbf h},{\boldsymbol \Omega})$ is
\begin{align}\label{MMSE}
(\hat{\mathbf z},\hat{\mathbf h},\hat{{\boldsymbol \Omega}})={\rm E}[({\mathbf z},{\mathbf h},{\boldsymbol \Omega})|\tilde {\mathbf y};{\boldsymbol \eta}_{\rm ML}],
\end{align}
where the expectation is taken with respect to
\begin{align}
p({\mathbf z},{\mathbf h},{\boldsymbol \Omega}|{\mathbf y };\hat{\boldsymbol \eta}_{\rm ML})=\frac{p({\mathbf z},{\mathbf h},{\boldsymbol \Omega},{\mathbf y };\hat{\boldsymbol \eta}_{\rm ML})}{p({\mathbf y };\hat{\boldsymbol \eta}_{\rm ML})}
\end{align}
Directly solving the ML estimate of $\boldsymbol \eta$ (\ref{MLbeta}) or the MMSE estimate of $({\mathbf z},{\mathbf h},{\boldsymbol \Omega})$ (\ref{MMSE}) are both intractable. As a result, an iterative algorithm is designed in Section \ref{Algorithm}.
\subsection{Module Design}\label{moddesign}
\begin{figure}[h!t]
\centering
\includegraphics[width=3.2in]{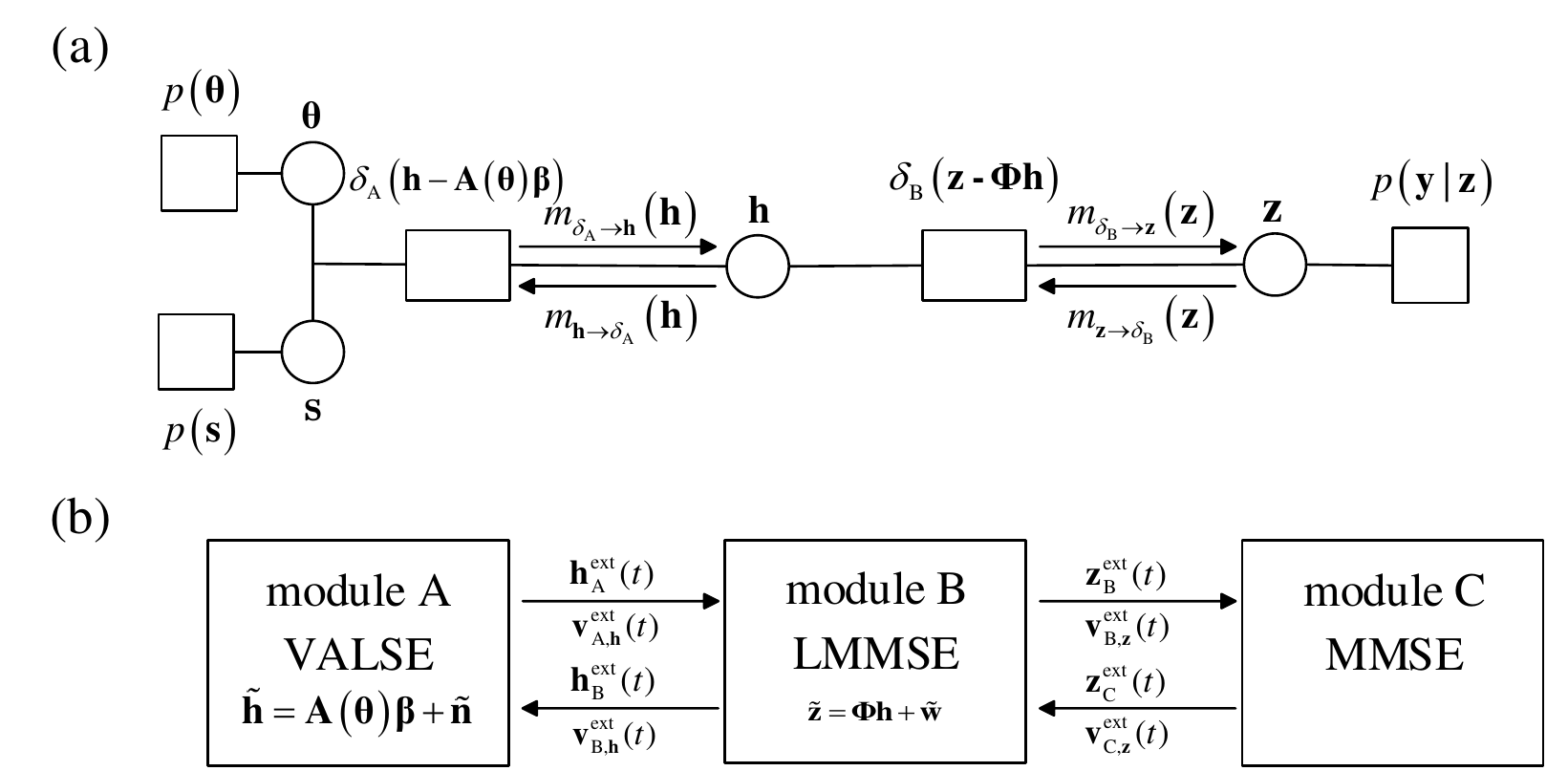}
\caption{Factor graph of the joint PDF (\ref{jointpdf}) and the module of the GL-QVBCE algorithm. Here the circles denote variable nodes, and the rectangles denote the factor node. According to the factor graphp in Fig. 1 (a), the problem can be decomposed as three modules in Fig. 1 (b), where module A corresponds to the standard LSE model, and module B corresponds to the standard linear model, module C corresponds to the MMSE estimation. Intuitively, the problem can be solved by iterating between the three modules, where module A performs the standard VALSE algorithm, module B performs the LMMSE, and module C performs the componentwise MMSE estimation.}
\label{FC_fig}                                                                                                                                                                                                                                                       \end{figure}

The additional hidden variables ${\mathbf z}$ defined in (\ref{defz}) and two $\delta(\cdot)$ factor nodes in the factor graph are the key to design GL-QVBCE algorithms. pAccording to the factor graph Fig. \ref{FC_fig} (a), the problem can be solved by exchanging information between each two modules, i.e., module A, module B and module C, where module A performs the variational line spectral estimation (VALSE) algorithm, module B performs the linear minimum mean square error estimation (LMMSE), and module C performs the componentwise minimum mean square error (MMSE) estimation \footnote{For the mixed one-bit system, the difference is that only the one-bit measurements are needed to be input to the MMSE module C, and the extrinsic information from module C and the unquantized measurements are input to the LMMSE module.}.

First, we initialize $m_{\delta_{\rm A}\rightarrow {\mathbf h}}^{0}({\mathbf h})$ and $m_{\delta_{\rm B}\rightarrow {\mathbf z}}^{0}({\mathbf z})$ referring to Fig. \ref{FC_fig}, where the subscript is for the index of iteration. For the $t$th iteration with $1\le t \le T_{\rm o}$, the algorithm is detailed as follows.
\subsubsection{MMSE module}
For the $t$th iteration, assume $m_{\delta_{\rm B}\rightarrow {\mathbf z}}^{t}({\mathbf z})={\mathcal {CN}}({\mathbf z};{\mathbf z}_{\rm B}^{\rm ext}(t);{\rm diag}({\mathbf v}_{\rm B,z}^{\rm ext}(t)))$. According to EP \cite{Minka}, the message $m_{{\mathbf z}\rightarrow\delta_{\rm B}}^{t}({\mathbf z})$ is calculated to be
\begin{align}\label{extCz}
m_{{\mathbf z}\rightarrow\delta_{\rm B}}^{t}({\mathbf z})\propto \frac{{\rm Proj}[m_{\delta_{\rm B}\rightarrow {\mathbf z}}^{t}({\mathbf z})p({\mathbf y}|{\mathbf z})]}{m_{\delta_{\rm B}\rightarrow {\mathbf z}}^{t}({\mathbf z})}\triangleq \frac{{\rm Proj}[q_{\rm C}^t({\mathbf z})]}{m_{\delta_{\rm B}\rightarrow {\mathbf z}}^{t}({\mathbf z})},
\end{align}
where $\propto $ denotes identity up to a normalizing constant. First, the MMSE estimate of $\mathbf z$ can be obtained, i.e.,
\begin{align}
&{\mathbf z}_{\rm C}^{\rm post}(t)={\rm E}[{\mathbf z}|q_{\rm C}^t({\mathbf z})],\label{comb_means}\\
&{\mathbf v}_{\rm C,z}^{\rm post}(t)={\rm Var}[{\mathbf z}|q_{\rm C}^t({\mathbf z})]\label{comb_vars},
\end{align}
where ${\rm E}[\cdot|q_{\rm C}^t({\mathbf z})]$ and ${\rm Var}[\cdot|q_{\rm C}^t({\mathbf z})]$ are the mean and variance operations taken componentwise with respect to the distribution $\propto q_{\rm C}^t({\mathbf z})$. As a result, ${\rm Proj}[q_{\rm C}^t({\mathbf z})]$ is \footnote{Here the diagonal EP is used. It is numerically shown that the scalar EP which averages the noise variance yields significant performance degradation.}
\begin{align}\label{postCz}
{\rm Proj}[q_{\rm C}^t({\mathbf z})]={\mathcal {CN}}({\mathbf z};{\mathbf z}_{\rm C}^{\rm post}(t),{\rm diag}({\mathbf v}_{\rm C,z}^{\rm post}(t))).
\end{align}
Substituting (\ref{postCz}) in (\ref{extCz}), the message $m_{{\mathbf z}\rightarrow \delta_{\rm B}}^t({\mathbf z})$ from the variable node $\mathbf z$ to the factor node $\delta_{\rm B}({\mathbf z}-{\boldsymbol \Phi}{\mathbf h})$ is calculated as
\begin{align}\label{extBtoA}
m_{{\mathbf z}\rightarrow \delta_{\rm B}}^t({\mathbf z})\propto \frac{{\mathcal {CN}}({\mathbf z};{\mathbf z}_{\rm C}^{\rm post}(t),{\rm diag( {\mathbf v}_{\rm C,z}^{\rm post}(t))})}{{\mathcal {CN}}({\mathbf z};{\mathbf z}_{\rm B}^{\rm ext}(t),{\rm diag ({\mathbf v}_{\rm B,z}^{\rm ext}(t))})}\propto {\mathcal {CN}}({\mathbf z};{\mathbf z}_{\rm C}^{\rm ext}(t),{\rm diag }({\mathbf v}_{\rm C,z}^{\rm ext}(t))),
\end{align}
where ${\mathbf z}_{\rm C}^{\rm ext}(t)$ and ${\mathbf v}_{\rm C,z}^{\rm ext}(t)$ are \cite{meng1}
\begin{subequations}
\begin{align}
&{\mathbf v}_{\rm C,z}^{\rm ext}(t)=\left(\frac{1}{{\mathbf v}_{\rm C,z}^{\rm post}(t)}-\frac{1}{{\mathbf v}_{\rm B,z}^{\rm ext}(t)}\right)^{-1},\label{extC_var}\\
&{\mathbf z}_{\rm C}^{\rm ext}(t)={\mathbf v}_{\rm C,z}^{\rm ext}(t)\odot\left(\frac{{\mathbf z}_{\rm C}^{\rm post}(t)}{{\mathbf v}_{\rm C,z}^{\rm post}(t)}-\frac{{\mathbf z}_{\rm B}^{\rm ext}(t)}{{\mathbf v}_{\rm B,z}^{\rm ext}(t)}\right),\label{extC_mean}
\end{align}
\end{subequations}
where $\odot$ denotes componentwise multiplication.
\subsubsection{LMMSE module}
According to the factor node $\delta_{\rm B}({\mathbf z}-{\boldsymbol \Phi}{\mathbf h})$ and the extrinsic message $m_{{\mathbf z}\rightarrow \delta_{\rm B}}^t({\mathbf z})$ transmitted from module C to module B, we obtain a pseudo standard linear model \footnote{For the multiuser model with equation (\ref{pseudo}), we extract the $k$th user information similarly as (\ref{extuser}) does and channel estimation can be performed.}
\begin{align}\label{pseudo}
\tilde{\mathbf z}={\boldsymbol \Phi}{\mathbf h}+\tilde{\mathbf w},
\end{align}
where $\tilde{\mathbf z}={\mathbf z}_{\rm C}^{\rm ext}(t)$, $\tilde{\mathbf w}\sim {\mathcal {CN}}(\tilde{\mathbf w};{\mathbf 0},{\rm diag}(\tilde{\boldsymbol \sigma}_w^2(t)))$, $\tilde{\boldsymbol \sigma}_w^2(t)={\mathbf v}_{\rm C,z}^{\rm ext}(t)$. Combing the message $m_{\delta_{\rm A}\rightarrow {\mathbf h}}^{t}({\mathbf h})={\mathcal {CN}}= {\mathcal {CN}}({\mathbf h};{\mathbf h}_{\rm A}^{\rm ext}(t),{\rm diag}({\mathbf v}_{\rm A,h}^{\rm ext}(t)))$, we update the message $m_{\delta_{\rm B}\rightarrow {\mathbf h}}^{t}({\mathbf h})$ as
\begin{align}
m_{\delta_{\rm B}\rightarrow {\mathbf h}}^{t}({\mathbf h})&=\frac{{\rm Proj}[\int m_{\delta_{\rm A}\rightarrow {\mathbf h}}^{t}({\mathbf h})\delta_{\rm B}({\mathbf z}-{\boldsymbol \Phi}{\mathbf h})m_{{\mathbf z}\rightarrow \delta_{\rm B}}^t({\mathbf z}){\rm d}{\mathbf z}
]}{m_{\delta_{\rm A}\rightarrow {\mathbf h}}^{t}({\mathbf h})}\notag\\
&\triangleq \frac{{\rm Proj}[q_{\rm B}^t({\mathbf h})]}{m_{\delta_{\rm A}\rightarrow {\mathbf h}}^{t}({\mathbf h})}.
\end{align}
First, it can be shown that $q_{\rm B}^t({\mathbf h})$ is Gaussian distributed with covariance matrix and mean being
\begin{align}
&{\boldsymbol \Sigma}_{\rm B,h}^{\rm post}=\left({\boldsymbol \Phi}^{\rm H}{\rm diag}(\tilde{\boldsymbol \sigma}_w^{-2}(t)){\boldsymbol \Phi}+{\rm diag}(1/{\mathbf v}_{\rm A,h}^{\rm ext}(t))\right)^{-1},\label{invlmmse1}\\
&{\mathbf h}_{\rm B}^{\rm post}={\boldsymbol \Sigma}_{\rm B,h}^{\rm post}\left({\boldsymbol \Phi}^{\rm H}{\rm diag}(\tilde{\boldsymbol \sigma}_w^{-2}(t))\tilde{\mathbf z}+{\mathbf h}_{\rm A}^{\rm ext}(t)/{\mathbf v}_{\rm A,h}^{\rm ext}(t)\right).
\end{align}
As a result, we project $q_{\rm B}^t({\mathbf h})$ as
\begin{align}
{\rm Proj}[q_{\rm B}^t({\mathbf h})]={\mathcal {CN}}({\mathbf h};{\mathbf h}_{\rm B}^{\rm post},{\rm diag}({\mathbf v}_{\rm B,h}^{\rm post})),
\end{align}
where
\begin{align}
{\mathbf v}_{\rm B,h}^{\rm post}={\rm diag}({\boldsymbol \Sigma}_{\rm B,h}^{\rm post}).
\end{align}
As a result, $m_{\delta_{\rm B}\rightarrow {\mathbf h}}^{t}({\mathbf h})$ can be calculated as
\begin{align}
m_{\delta_{\rm B}\rightarrow {\mathbf h}}^{t}({\mathbf h})&\propto \frac{{\mathcal {CN}}({\mathbf h};{\mathbf h}_{\rm B}^{\rm post}(t),{\rm diag( {\mathbf v}_{\rm B,h}^{\rm post}(t))})}{{\mathcal {CN}}({\mathbf h};{\mathbf h}_{\rm A}^{\rm ext}(t),{\rm diag ({\mathbf v}_{\rm A,h}^{\rm ext}(t))})}\propto {\mathcal {CN}}({\mathbf h};{\mathbf h}_{\rm B}^{\rm ext}(t),{\rm diag }({\mathbf v}_{\rm B,h}^{\rm ext}(t))),
\end{align}
where
\begin{subequations}
\begin{align}
&{\mathbf v}_{\rm B,h}^{\rm ext}(t)=\left(\frac{1}{{\mathbf v}_{\rm B.h}^{\rm post}(t)}-\frac{1}{{\mathbf v}_{\rm A,h}^{\rm ext}(t)}\right)^{-1},\label{extB_var}\\
&{\mathbf h}_{\rm B}^{\rm ext}(t)={\mathbf v}_{\rm B,h}^{\rm ext}(t)\odot\left(\frac{{\mathbf h}_{\rm B}^{\rm post}(t)}{{\mathbf v}_{\rm B,h}^{\rm post}(t)}-\frac{{\mathbf h}_{\rm A}^{\rm ext}(t)}{{\mathbf v}_{\rm A,h}^{\rm ext}(t)}\right),\label{extB_mean}
\end{align}
\end{subequations}
\subsubsection{VALSE module}
According to $m_{\delta_{\rm B}\rightarrow {\mathbf h}}^{t}({\mathbf h})$, we obtain another pseudo measurement model
\begin{align}
\tilde{\mathbf h}={\mathbf A}({\boldsymbol \theta}){\boldsymbol \beta}+\tilde{\mathbf n},
\end{align}
where $\tilde{\mathbf h}={\mathbf h}_{\rm B}^{\rm ext}(t)$, $\tilde{\mathbf n}\sim {\mathcal {CN}}(\tilde{\mathbf n};{\mathbf 0},{\rm diag}({\mathbf v}_{\rm B,h}^{\rm ext}(t)))$. Now we run the VALSE algorithm \cite{JiangVALSEQ} and calculate the posterior mean and variances of ${\mathbf h}$ as ${\mathbf h}_{\rm A}^{\rm post}(t)$ and ${\mathbf v}_{\rm A,h}^{\rm post}(t)$, respectively.
\subsubsection{From VALSE to LMMSE}
We now update the message $m_{\delta_{\rm A}\rightarrow {\mathbf h}}^{t+1}({\mathbf h})={\mathcal {CN}}({\mathbf h};{\mathbf h}_{\rm A}^{\rm ext}(t+1),{\rm diag}({\mathbf v}_{\rm A,h}^{\rm ext}(t+1)))$, given by
\begin{align}
&{\mathbf v}_{\rm A,h}^{\rm ext}(t+1)=\left(\frac{1}{{\mathbf v}_{\rm A,h}^{\rm post}(t)}-\frac{1}{{\mathbf v}_{\rm B,h}^{\rm ext}(t)}\right)^{-1},\\
&{\mathbf h}_{\rm A}^{\rm ext}(t+1)={\mathbf v}_{\rm A,h}^{\rm ext}(t+1)\odot\left(\frac{{\mathbf h}_{\rm A}^{\rm post}(t)}{{\mathbf v}_{\rm A,h}^{\rm post}(t)}-\frac{{\mathbf h}_{\rm B}^{\rm ext}(t)}{{\mathbf v}_{\rm B,h}^{\rm ext}(t)}\right).
\end{align}
\subsubsection{From LMMSE to MMSE}
We pupdate the message $m_{\delta_{\rm B}\rightarrow {\mathbf z}}^{t+1}({\mathbf z})={\mathcal {CN}}({\mathbf z};{\mathbf z}_{\rm B}^{\rm ext}(t+1);{\rm diag}({\mathbf v}_{\rm B}^{\rm ext}(t+1)))$. It can be seen that once this message is updated, the iteration is closed. The message is updated as follows:
\begin{align}\label{zBextequ}
m_{\delta_{\rm B}\rightarrow {\mathbf z}}^{t+1}({\mathbf z})&=\frac{{\rm Proj}[\int m_{\delta_{\rm A}\rightarrow {\mathbf h}}^{t+1}({\mathbf h})\delta_{\rm B}({\mathbf z}-{\boldsymbol \Phi}{\mathbf h})m_{{\mathbf z}\rightarrow \delta_{\rm B}}^{t}({\mathbf z}){\rm d}{\mathbf h}
]}{m_{{\mathbf z}\rightarrow \delta_{\rm B}}^{t}({\mathbf z})}\triangleq \frac{{\rm Proj}[q_{\rm B}^{t+1}({\mathbf z})]}{m_{{\mathbf z}\rightarrow \delta_{\rm B}}^{t}({\mathbf z})}.
\end{align}
It can be calculated that
\begin{align}
q_{\rm B}^{t+1}({\mathbf z})={\mathcal {CN}}({\mathbf z};{\mathbf z}_{\rm B}^{\rm post}(t+1),{\boldsymbol \Sigma}_{\rm B}^{\rm post}(t+1)),
\end{align}
where
\begin{align}
&{\boldsymbol \Sigma}_{\rm B}^{\rm post}(t+1)={\boldsymbol \Phi}\left({\boldsymbol \Phi}^{\rm H}{\rm diag}(\tilde{\boldsymbol \sigma}_w^{-2}(t)){\boldsymbol \Phi}+{\rm diag}(1/{\mathbf v}_{\rm A,h}^{\rm ext}(t+1))\right)^{-1}{\boldsymbol \Phi}^{\rm H},\label{invlmmse2}\\
&{\mathbf z}_{\rm B}^{\rm post}(t+1)={\boldsymbol \Phi}\left({\boldsymbol \Phi}^{\rm H}{\rm diag}(\tilde{\boldsymbol \sigma}_w^{-2}(t)){\boldsymbol \Phi}+{\rm diag}(1/{\mathbf v}_{\rm A,h}^{\rm ext}(t+1))\right)^{-1}\notag\\
&\times \left({\boldsymbol \Phi}^{\rm H}{\rm diag}(\tilde{\boldsymbol \sigma}_w^{-2}(t))\tilde{\mathbf z}+{\mathbf h}_{\rm A}^{\rm ext}(t+1)/{\mathbf v}_{\rm A,h}^{\rm ext}(t+1)\right).
\end{align}
Let ${\mathbf v}_{\rm B}^{\rm post}(t+1)={\rm diag}({\boldsymbol \Sigma}_{\rm B}^{\rm post}(t+1))$. Then
\begin{align}
{\rm Proj}[q_{\rm B}^{t+1}({\mathbf z})]={\mathcal {CN}}({\mathbf z};{\mathbf z}_{\rm B}^{\rm post}(t+1),{\rm diag}({\mathbf v}_{\rm B}^{\rm post}(t+1))).
\end{align}
Utilizing (\ref{zBextequ}), one has
\begin{subequations}
\begin{align}
&{\mathbf v}_{\rm B,z}^{\rm ext}(t+1)=\left(\frac{1}{{\mathbf v}_{\rm B,z}^{\rm post}(t+1)}-\frac{1}{{\mathbf v}_{\rm B,z}^{\rm ext}(t)}\right)^{-1},\\
&{\mathbf z}_{\rm B}^{\rm ext}(t+1)={\mathbf v}_{\rm B,z}^{\rm ext}(t+1)\odot\left(\frac{{\mathbf z}_{\rm B}^{\rm post}(t+1)}{{\mathbf v}_{\rm B,z}^{\rm post}(t+1)}-\frac{{\mathbf z}_{\rm B}^{\rm ext}(t)}{{\mathbf v}_{\rm B,z}^{\rm ext}(t)}\right),
\end{align}
\end{subequations}
which closes the loop of the proposed Gridless quantized variational Bayesian channel estimation (GL-QVBCE) algorithm.

The computation complexity of the proposed algorithm is analyzed for a single outer iteration. For the VALSE module, its computation complexity is $O(MN+NL^3)$ \cite{JiangVALSEQ}. While for the LMMSE module, the computations are dominated by (\ref{invlmmse1}) and (\ref{invlmmse2}) involving a $M\times M$ matrix inversion, whose computation complexity is $O(M^3)$. As for the componentwise MMSE module, its computation complexity is small. The overall computation complexity is $O(M^3+MN+NL^3)$. As a comparison, the atomic norm based algorithm performing cvx \cite{cvx} has a complexity of $O(M^{4}N^{2.5}+N^{4.5})$ \cite{Yang1}. Since the number of iterations is usually small for the GL-QVBCE, the computation complexity of atomic norm based algorithm is higher than that of the proposed approach.
\subsection{Extension to Sequential Channel Estimation}
In some settings, the DOAs are fixed across the pilots, while the gains and phases are varied. As a result, performing the channel tracking is very important. Similar to \cite{JiangVALSE}, sequential GL-QVBCE (Seq-GL-QVBCE) is developed. Since GL-QVBCE outputs the posterior PDF of the frequencies, performing the sequential estimation is very natural.

Suppose that we have performed GL-QVBCE given the first pilot $x_t|_{t=1}$ and obtain the posterior PDF $p({\boldsymbol \theta}_{t}|{t=1})$ of $\boldsymbol \theta$. For the second pilot $x_t|_{t=2}$, we run the GL-QVBCE algorithm. For the VALSE module, we encode $p({\boldsymbol \theta}_{t}|{t=1})$ as the prior distribution of $\boldsymbol \theta$. Following the previous steps, Seq-GL-QVBCE can be developed.

The prior distribution obtained from the previous measurements may be too strong and inaccurate, which will deteriorate the estimation performance. As a result, we may use the damping operation to decrease the concentration parameter ${\boldsymbol \kappa}_{0,t}\in {\mathbb R}^L$ of the prior distribution $p({\boldsymbol \theta}_{t})$ \footnote{For large $\kappa$, the variance of the frequency is approximately $1/\kappa$.}. For the $(t-1)$th posterior PDF $p({\boldsymbol \theta}_{t-1})$ with concentration parameter ${\boldsymbol \kappa}_{t-1}$, we set
\begin{align}\label{lambdaset}
{\boldsymbol \kappa}_{0,t}=\lambda{\boldsymbol \kappa}_{t-1},
\end{align}
where $0<\lambda\leq 1$.
\section{Cram\'{e}r Rao bound}\label{bound}
Before designing the recovery algorithm, the performance bounds of unbiased estimators are derived, i.e., the Cram\'{e}r Rao bound (CRB). Although the Bayesian algorithm is designed, the CRB acts as the performance benchmark of the algorithm. To derive the CRB, $L$ is assumed to be known, the frequencies ${\boldsymbol \theta}\in{\mathbb R}^L$ (or DOAs ${\boldsymbol \phi}\in{\mathbb R}^L$) and weights ${\boldsymbol \beta}\in{\mathbb C}^L$ are treated as deterministic unknown parameters. As for the quantizer $Q(\cdot)$, the quantization intervals are $\{(t_b,t_{b+1})\}_{b=0}^{|{\mathcal D}|-1}$, where $t_0=-\infty$, $t_{{\mathcal D}}=\infty$, $\bigcup_{b=0}^{{\mathcal D}-1}[t_b,t_{b+1})={\mathbb R}$. Given a real number $a\in [t_b,t_{b+1})$, the representation is
\begin{align}
Q(a)=\omega_b, \quad {\rm if}\quad a\in [t_b,t_{b+1}).
\end{align}
Note that for a quantizer with bit-depth $B$, the cardinality of the output of the quantizer is $|{\mathcal D}|=2^B$. Let $\boldsymbol \kappa$ denote the set of parameters, i.e., ${\boldsymbol \kappa}=[{\boldsymbol \theta}^{\rm T},{\mathbf g}^{\rm T},{\boldsymbol \varphi}^{\rm T}]^{\rm T}\in {\mathbb R}^{3L}$, where ${\boldsymbol \beta}={\mathbf g}{\rm e}^{{\rm j}{\boldsymbol \varphi}}$. The probability mass function (PMF) of the measurements $p({\mathbf Y}|{\boldsymbol \kappa})$ is
\begin{align}
p({\mathbf Y}|{\boldsymbol \kappa})=\prod\limits_{i=1}^M\prod\limits_{t=1}^Tp(Y_{it}|{\boldsymbol \kappa})=\prod\limits_{i=1}^M\prod\limits_{t=1}^Tp(\Re\{Y_{it}\}|{\boldsymbol \kappa})p(\Im\{Y_{it}\}|{\boldsymbol \kappa}).
\end{align}
Moreover, the PMFs of $\Re\{Y_{it}\}$ and $\Im\{Y_{it}\}$ are
\begin{align}
p(\Re\{Y_{it}\}|{\boldsymbol \kappa})=\prod\limits_{\omega_b\in {\mathcal D}}p_{\Re\{Y_{it}\}}(\omega_b|{\boldsymbol \kappa})^{I_{Q(\Re\{Y_{it}\})=\omega_b}},\\
p(\Im\{Y_{it}\}|{\boldsymbol \kappa})=\prod\limits_{\omega_b\in {\mathcal D}}p_{\Im\{Y_{it}\}}(\omega_b|{\boldsymbol \kappa})^{I_{Q(\Im\{Y_{it}\})=\omega_b}},
\end{align}
where $I_{(\cdot)}$ is the indicator function,
\begin{align}
&p_{\Re\{Y_{it}\}}(\omega_l|{\boldsymbol \kappa})={\rm P}\left(\Re\{Y_{it}\}\in[t_b,t_{b+1})\right)={\Phi(\frac{t_{b+1}-\Re\{Z_{it}\}}{\sigma/\sqrt{2}})
-\Phi(\frac{t_{b}-\Re\{Z_{it}\}}{\sigma/\sqrt{2}})},\\
&p_{\Im\{Y_{it}\}}(\omega_l|{\boldsymbol \kappa})={\rm P}\left(\Im\{Y_{it}\}\in[t_b,t_{b+1})\right)
={\Phi(\frac{t_{b+1}-\Im\{Z_{it}\}}{\sigma/\sqrt{2}})-\Phi(\frac{t_{b}-\Im\{Z_{it}\}}{\sigma/\sqrt{2}})}.
\end{align}
The CRB is equal to the inverse of the Fisher information matrix (FIM) ${\mathbf I}({\boldsymbol \kappa})\in{\mathbb R}^{3L\times 3L}$
\begin{align}
{\mathbf I}({\boldsymbol \kappa})={\rm E}\left[\left(\frac{\partial \log p({\mathbf y}|{\boldsymbol \kappa})}{\partial {\boldsymbol \kappa}}\right)\left(\frac{\partial \log p({\mathbf y}|{\boldsymbol \kappa})}{\partial {\boldsymbol \kappa}}\right)^{\rm T}\right].
\end{align}
To calculate the FIM, the following Theorem \cite{Fu} is utilized.
\begin{theorem}\label{FIMlemma}
\cite{Fu} The FIM ${\mathbf I}({\boldsymbol \kappa})$ for estimating the unknown parameter ${\boldsymbol \kappa}$ is
\begin{align}\label{quanFIM}
{\mathbf I}({\boldsymbol \kappa})=\sum\limits_{i=1}^M\sum\limits_{t=1}^T   \left({\lambda}_i\frac{\partial \Re\{Z_{it}\}}{\partial {\boldsymbol \kappa}}\left(\frac{\partial \Re\{Z_{it}\}}{\partial {\boldsymbol \kappa}}\right)^{\rm T}+\chi_i\frac{\partial \Im\{Z_{it}\}}{\partial {\boldsymbol \kappa}}\left(\frac{\partial \Im\{Z_{it}\}   }{\partial {\boldsymbol \kappa}}\right)^{\rm T}\right).
\end{align}
For a general quantizer, one has
\begin{align}\label{lambdai}
\lambda_i=\frac{2}{\sigma^2}\sum\limits_{l=0}^{|{\mathcal D}|-1}\frac{[\phi(\frac{t_{b+1}-\Re\{Z_{it}\}}{\sigma/\sqrt{2}})-\phi(\frac{t_{b}-\Re\{Z_{it}\}}{\sigma/\sqrt{2}})]^2}{\Phi(\frac{t_{b+1}-\Re\{Z_{it}\}}{\sigma/\sqrt{2}})-\Phi(\frac{t_{b}-\Re\{Z_{it}\}}{\sigma/\sqrt{2}})},
\end{align}
and
\begin{align}\label{chii}
\chi_i=\frac{2}{\sigma^2}\sum\limits_{l=0}^{|{\mathcal D}|-1}\frac{[\phi(\frac{t_{b+1}-\Im\{Z_{it}\}}{\sigma/\sqrt{2}})-\phi(\frac{t_{b}-\Im\{Z_{it}\}}{\sigma/\sqrt{2}})]^2}{\Phi(\frac{t_{b+1}-\Im\{Z_{it}\}}{\sigma/\sqrt{2}})-\Phi(\frac{t_{b}-\Im\{Z_{it}\}}{\sigma/\sqrt{2}})},
\end{align}
For the unquantized system, the FIM is
\begin{align}\label{unqFIM}
&{\mathbf I}_{\rm unq}({\boldsymbol \kappa})=\frac{2}{\sigma^2}\sum\limits_{i=1}^M \sum\limits_{t=1}^T  \left(\frac{\partial \Re\{Z_{it}\}}{\partial {\boldsymbol \kappa}}\left(\frac{\partial \Re\{Z_{it}\}}{\partial {\boldsymbol \kappa}}\right)^{\rm T}+\frac{\partial \Im\{Z_{it}\}}{\partial {\boldsymbol \kappa}}\left(\frac{\partial \Im\{Z_{it}\}   }{\partial {\boldsymbol \kappa}}\right)^{\rm T}\right)
\end{align}
\end{theorem}

According to Theorem \ref{FIMlemma}, we need to calculate $\frac{\partial \Re\{Z_{it}\}}{\partial {\boldsymbol \kappa}}$ and $\frac{\partial \Im\{Z_{it}\}}{\partial {\boldsymbol \kappa}}$.
In our setting, we have
\begin{align}
\Re\{{\mathbf Z}\}&=[\Re\{{\mathbf A}\}\Re\{{\boldsymbol \beta}\}-\Im\{{\mathbf A}\}\Im\{{\boldsymbol \beta}\}]\Re\{{\mathbf x}^{\rm T}\}\notag\\
&-[\Re\{{\mathbf A}\}\Im\{{\boldsymbol \beta}\}+\Im\{{\mathbf A}\}\Re\{{\boldsymbol \beta}\}]\Im\{{\mathbf x}^{\rm T}\},\\
\Im\{{\mathbf Z}\}&=[\Re\{{\mathbf A}\}\Re\{\boldsymbol \beta\}-\Im\{{\mathbf A}\}\Im\{{\boldsymbol \beta}\}]\Im\{{\mathbf x}^{\rm T}\}\notag\\
&+[\Re\{{\mathbf A}\}\Im\{{\boldsymbol \beta}\}+\Im\{{\mathbf A}\}\Re\{{\boldsymbol \beta}\}]\Re\{{\mathbf x}^{\rm T}\}.
\end{align}
We need to calculate $\frac{\partial \Re\{Z_{it}\}}{\partial {\boldsymbol \kappa}}$ and $\frac{\partial \Im\{Z_{it}\}}{\partial {\boldsymbol \kappa}}$. By defining
\begin{align}
\xi_{il}=\sin(m_i\theta_l)\sin(\varphi_l)-\cos(m_i\theta_l)\cos(\varphi_l),\\
\zeta_{il}=\sin(m_i\theta_l)\cos(\varphi_l)+\cos(m_i\theta_l)\sin(\varphi_l).
\end{align}

We have, for $l=1,\cdots,L$,
\begin{subequations}\label{partialder}
\begin{align}
&\frac{\partial \Re\{Z_{it}\}}{\partial {\theta_l}}=m_ig_l(\xi_{il}\Im\{{x}_t\}-\zeta_{il}\Re\{{x}_t\})
\\
%&\frac{\partial \theta_l}{\partial {\phi_l}}=\frac{\pi^2}{180}\cos(\frac{\pi}{180}\phi)\\
&\frac{\partial \Re\{Z_{it}\}}{\partial {g_l}}=-\xi_{il}\Re\{{x}_t\}-\zeta_{il}\Im\{{x}_t\}\\
&\frac{\partial \Re\{Z_{it}\}}{\partial {\varphi_l}}=-g_l(\zeta_{il}\Re\{{x}_t\}-\xi_{il}\Im\{{x}_t\})\\
&\frac{\partial \Im\{Z_{it}\}}{\partial {\theta_l}}=m_ig_l(-\xi_{il}\Re\{{x}_t\}-\zeta_{il}\Im\{{x}_t\})
\\
&\frac{\partial \Im\{Z_{it}\}}{\partial {g_l}}=\zeta_{il}\Re\{{x}_t\}-\xi_{il}\Im\{{x}_t\}\\
&\frac{\partial \Im\{Z_{it}\}}{\partial {\varphi_l}}=-g_l(\xi_{il}\Re\{{x}_t\}+\zeta_{il}\Im\{{x}_t\}).
\end{align}
\end{subequations}
Substituting (\ref{partialder}), (\ref{lambdai}) and (\ref{chii}) in (\ref{quanFIM}) and (\ref{unqFIM}), respectively, one obtains the FIM for quantized and unquantized settings. The CRBs are the inverse of the corresponding FIMs, i.e., ${\rm CRB}({\boldsymbol \kappa})={\mathbf I}^{-1}({\boldsymbol \kappa})$ and ${\rm CRB}_{\rm unq}({\boldsymbol \kappa})={\mathbf I}_{\rm unq}^{-1}({\boldsymbol \kappa})$, respectively. The CRB of the frequencies, gains and phases are $[{\rm CRB}({\boldsymbol \kappa})]_{1:K,1:K}$, $[{\rm CRB}({\boldsymbol \kappa})]_{K+1:2K,K+1:2K}$, $[{\rm CRB}({\boldsymbol \kappa})]_{2K+1:3K,2K+1:3K}$, respectively, which will be used as the performance metrics.
\section{Numerical Simulation}
Performance of the proposed GL-QVBCE is evaluated numerically. The simulation setup is similar to \cite{Wen}. For the ULA of the channel vector ${\mathbf h}={\mathbf A}({\boldsymbol \phi}){\boldsymbol \beta}$, we generate the multipath channel including a line-of-sight (LoS) path and $L-1$ non-LoS paths. Let $P$ denote the total received power and $P=\sum\limits_{l=1}^L{\rm E}[|\beta_l|^2]$. For each run, the ray gains  $\{g_l\}_{l=1}^L$ are generated from the complex normal distribution, where ${\boldsymbol \beta}={\mathbf g}{\rm e}^{{\rm j}{\boldsymbol \varphi}}$. In addition, the average powers of the LoS ray gains are set as ${\rm E}[g_1^2]=0.5P$ and $g_1$ is generated from ${\mathcal N}(\sqrt{0.45P},0.05P)$, whereas the remaining non-LoS paths gains are ${\rm E}[g_l^2]=0.5P/(L-1)$, $l=2,\cdots,L$ and are generated from ${\mathcal N}(\sqrt{0.45P/(L-1)},0.05P/(L-1))$. The phases $\boldsymbol \varphi$ are generated uniformly from $(-\pi,\pi)$. The DOAs $\{\phi_l\}_{l=1}^L$ are chosen uniformly from $[-\pi/2,\pi/2)$. All the results are averaged over $100$ Monte Carlo (MC) trials unless stated otherwise. The signal to noise ratio (SNR) is defined as
\begin{align}
{\rm SNR}=\frac{{\rm E}[\|{\mathbf h}{\mathbf x}^{\rm T}\|_{\rm F}^2]}{{\rm E}[\|{\mathbf W}\|_{\rm F}^2]}=\frac{P}{\sigma^2}.
\end{align}

As for the quantizer, zero threshold is chosen for 1-bit quantizer, while a uniform quantizer is chosen for multi-bit quantization. For ${\mathbf Z}={\mathbf h}{\mathbf x}^{\rm T}$, straightforward calculation shows that the variance $\sigma_z^2$ of the elements of $\mathbf Z$ is $\sigma_z^2=P$. The real and imaginary parts of $\mathbf Z$ are quantized separately, and the dynamic range of the quantizer is restricted to be $[-3\sigma_z/\sqrt{2},3\sigma_z/\sqrt{2}]$. For a uniform quantizer with bit-depth $B$, the quantizer step size $\Delta$ is $\Delta=3\sigma_z/2^{B-0.5}$. Then the variance of the additive quantization noise is $\sigma_q^2=2\Delta^2/12=3\sigma_z^2/2^{2B}$ (including the real and imaginary parts) \cite{CSquant}. It is assumed that the noise variance $\sigma^2$ is available.

Three additional algorithms are implemented to make performance comparison. The first is the proposed GL-VBCE which works under unquantized setting. The second is GL-VBCE-AQNM which uses the AQNM model
\begin{align}\label{aqnmmodel}
{\mathbf Y}={\mathbf h}{\mathbf x}^{\rm T}+{\mathbf W}+{\mathbf N}_q\triangleq {\mathbf h}{\mathbf x}^{\rm T}+{\mathbf W}_{\rm eq},
\end{align}
where ${\mathbf N}_q$ denotes the additive quantization noise whose elements are i.i.d. and satisfies $[{\mathbf N}_q]_{i,t}={\mathcal {CN}}([{\mathbf N}_q]_{i,t};0,\sigma_q^2)$. Consequently, the AQNM model is the same as that of the unquantized model, and we apply the GL-VBCE with the equivalent noise ${\mathbf W}_{\rm eq}$, whose elements are i.i.d. and satisfies $[{\mathbf W}_{\rm eq}]_{i,t}={\mathcal {CN}}([{\mathbf W}_{\rm eq}]_{i,t};0,\sigma_w^2+\sigma_q^2)$. The last is the conventional LS approach works under unquantized setting. It can be easily shown that the NMSE of LS is
\begin{align}\label{LSperb}
{\rm NMSE}({\rm LS})=-{\rm SNR}-10\log T \quad ({\rm dB}),
\end{align}
i.e., only SNR or the number of pilots can improve the performance of LS. This phenomenon can also be validated by the ensuing numerical simulations.

\subsection{Channel Estimation Error versus Iteration}\label{CEvsiter}
\begin{figure}
\centering{\includegraphics[width=2.8in]{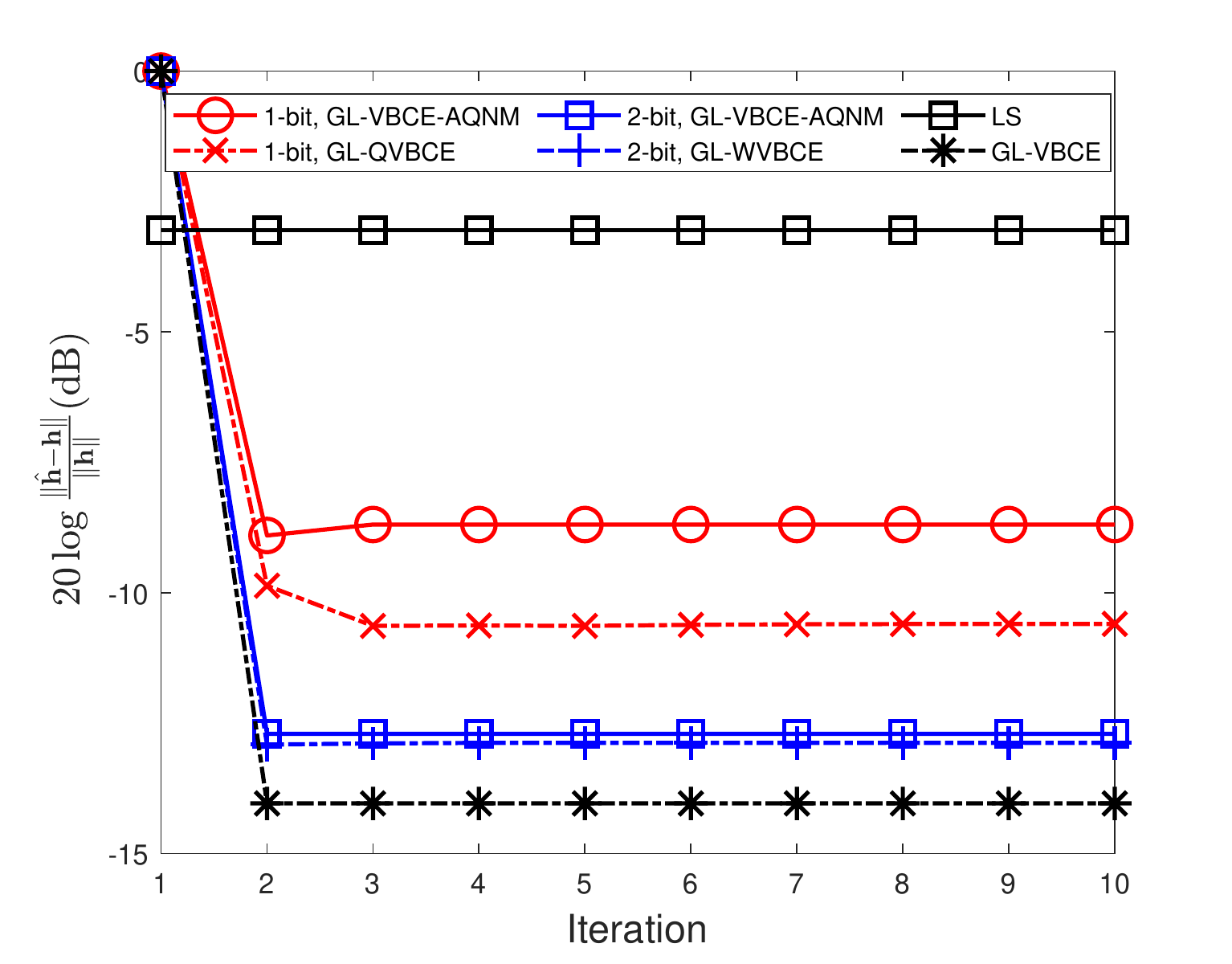}}
\caption{The mean NMSE of the channel versus the number of iterations for various algorithms averaged over $50$ MC trials. Here $N=M=48$, ${\rm SNR}=0$ dB, $T=2$.
}\label{MSE_iter}
\end{figure}

At first, the NMSEs of the algorithms versus the iteration are presented and shown in Fig. \ref{MSE_iter}. It can be seen that all the algorithms converge in several iterations. From (\ref{LSperb}), the NMSE of LS is $-10\log 2\approx -3$ dB due to ${\rm SNR}=0$ dB and $T=2$. All the algorithms utilizing channel structure works better than the traditional LS approach. Under 1-bit quantization, the performance of GL-QVBCE is better than that of GL-VBCE-AQNM. While for 2-bit quantization, the performance of GL-QVBCE is comparable to that of GL-VBCE-AQNM, and is close to GL-VBCE. Since GL-QVBCE involves the nonlinear MMSE, GL-VBCE-AQNM is preferable under 2-bit quantization in this setting.
\subsection{Comparisons of Channel Estimators versus SNR}\label{vsSNR}
\begin{figure}
\centering{\includegraphics[width=2.8in]{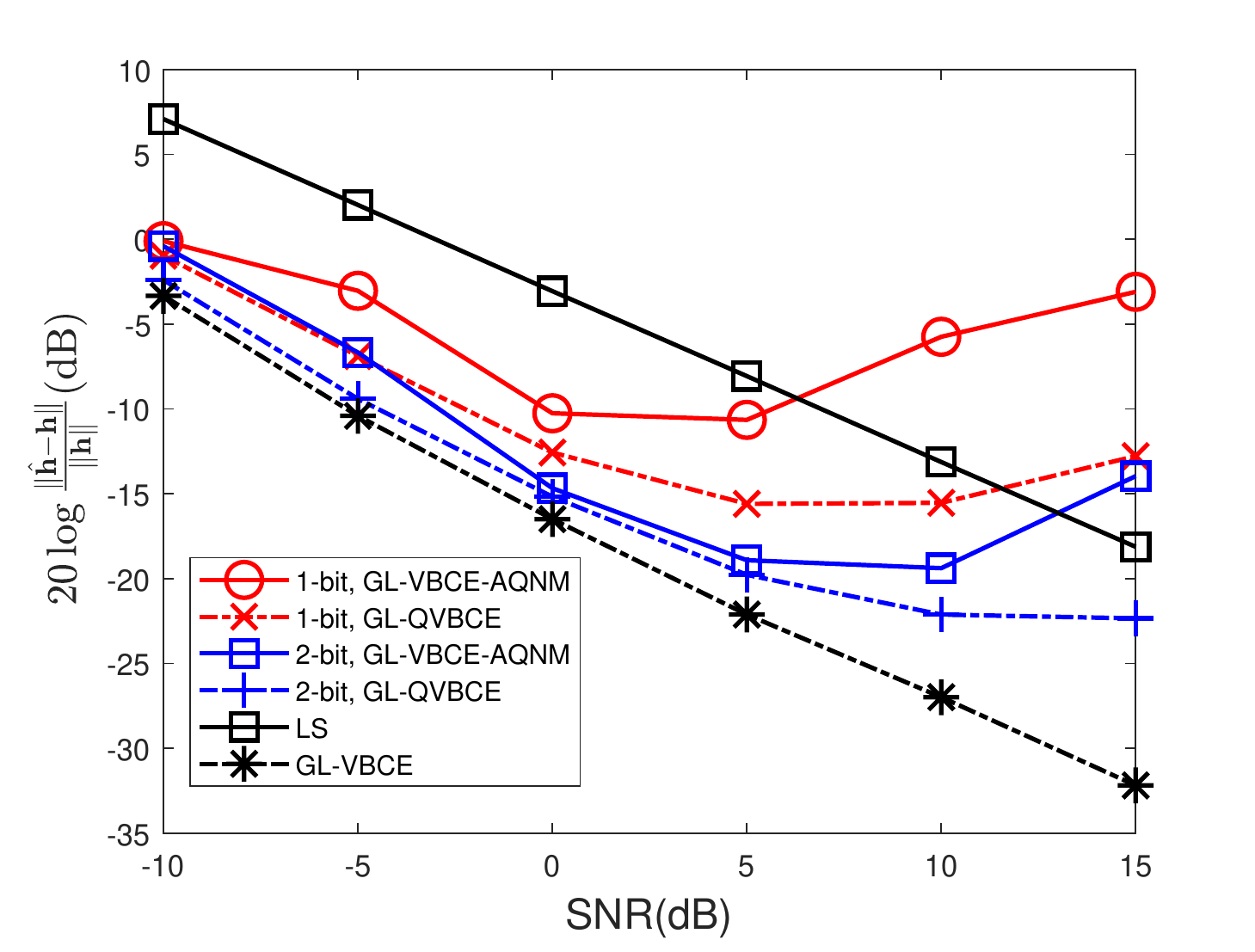}}
\caption{The mean NMSE of the channel versus SNR. Here $N=M=64$, $T=2$.
}\label{MSE_SNR}
\end{figure}

This subsection investigates the channel estimation performance versus SNR. Parameters are set as follows: $N=M=64$, $L=2$, $T=2$. The performance of channel estimation error is presented in Fig. \ref{MSE_SNR}. The NMSEs of GL-VBCE and LS decreases linearly with $\rm SNR$. For GL-QVBCE under two bit quantization, the NMSE decreases as SNR increases. When SNR increases to $10$ dB, the NMSE is saturated. For the GL-VBCE-AQNM and GL-QVBCE under one bit quantization, the NMSEs first decrease as SNR increases. When the SNR exceeds a certain value, the NMSEs begin to increase. The phenomenon from GL-VBCE-AQNM demonstrates that AQNM is more preferable when SNR is low.
\subsection{Comparisons with CRB}
\begin{figure}[h!t]
\centering{\includegraphics[width=2.8in]{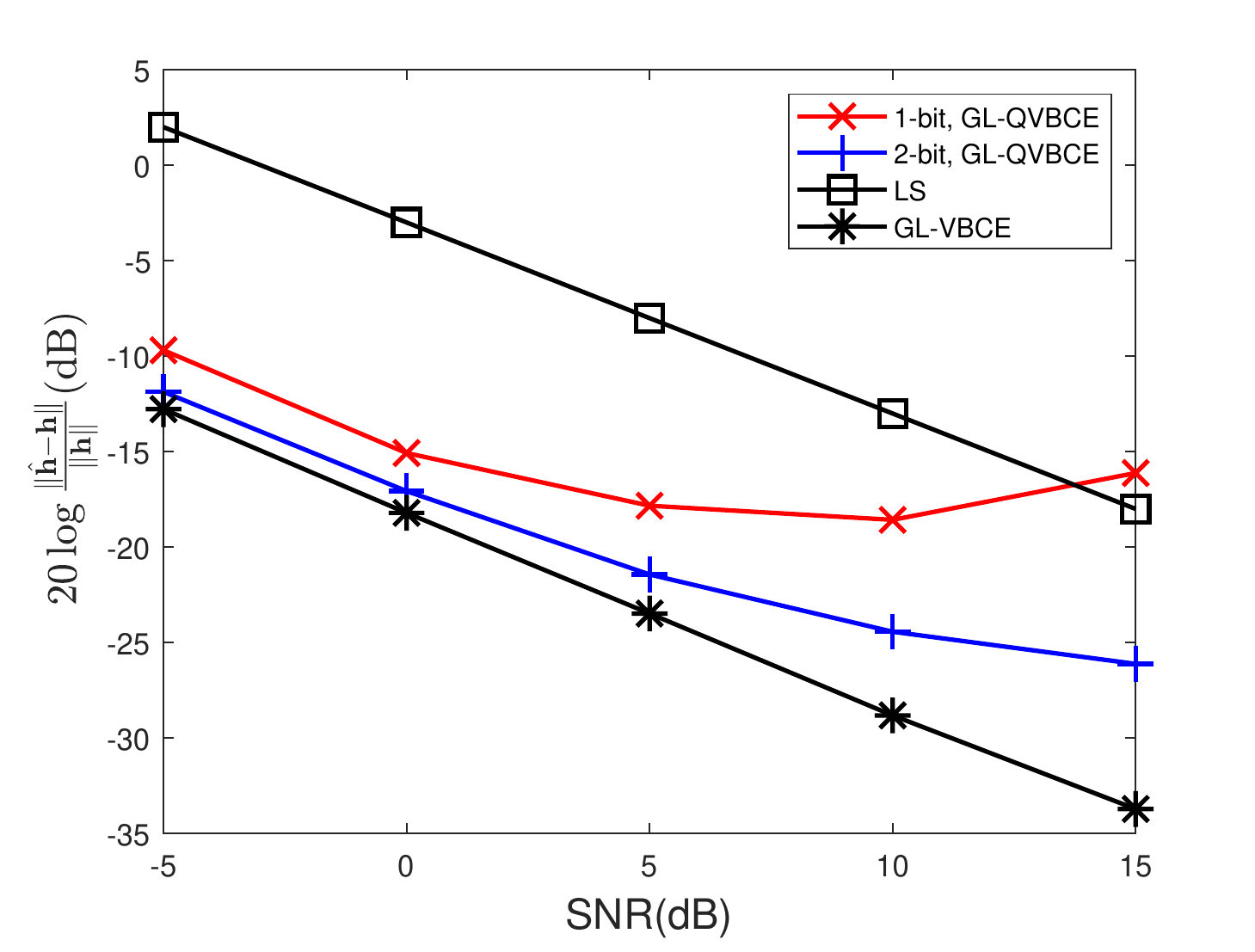}}
\caption{NMSEs of channel versus SNRs of the various estimators averaged over $500$ MC trials.
}\label{MSEchannel}
\end{figure}

The performances of the various algorithms against the CRB are investigated. Here we fix the two DOAs as $\phi=[-30^{\circ},60^{\circ}]^{\rm T}$, set the corresponding complex amplitudes as ${\boldsymbol \beta}=[0.8{\rm e}^{-{\rm j}0.3\pi},0.6{\rm e}^{{\rm j}0.2\pi}]^{\rm T}$. The pilot length is $T=2$. $N=M=96$. Here $P(\hat{L}=L)$ denotes the probability of both the model order $K$ is correctly estimated and ${\rm NMSE}(\hat{\mathbf h})\leq -5{\rm dB}$. As for the MSEs of ray-gain estimation, ray-phase estimation and DOA estimation, we only average the case when the two conditions are satisfied. For clarity, the performances of GL-VBCE-AQNM are not presented here. Fig. \ref{MSEchannel} presents the NMSEs versus SNR under this setting. The phenomena are basically the same as shown in subsection \ref{vsSNR}.

\begin{figure*}
  \centering
  \subfigure[]{
    \label{MSEg} %% label for first subfigure
    \includegraphics[width=2.8in]{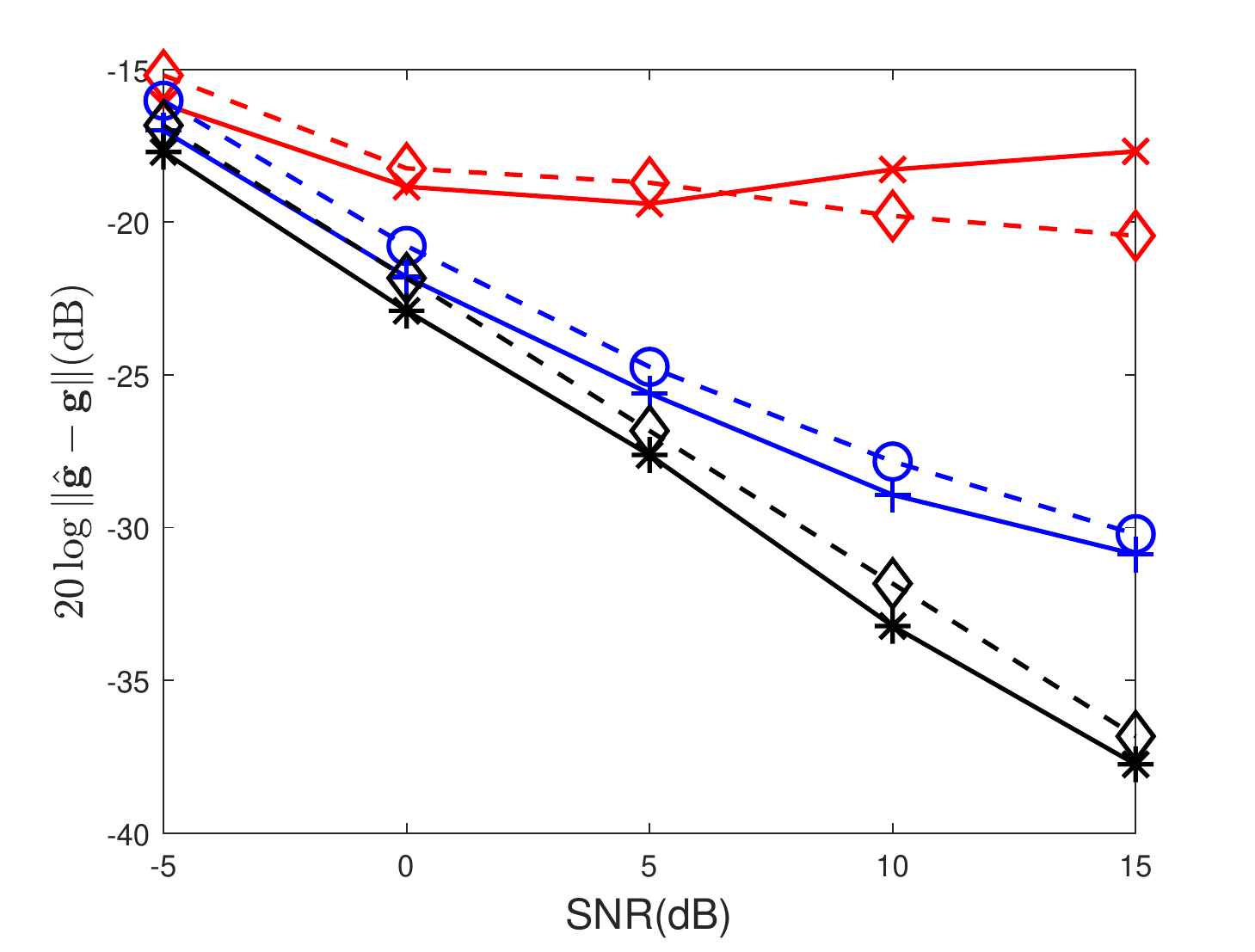}}
  \subfigure[]{
    \label{MSEphase} %% label for first subfigure
    \includegraphics[width=2.8in]{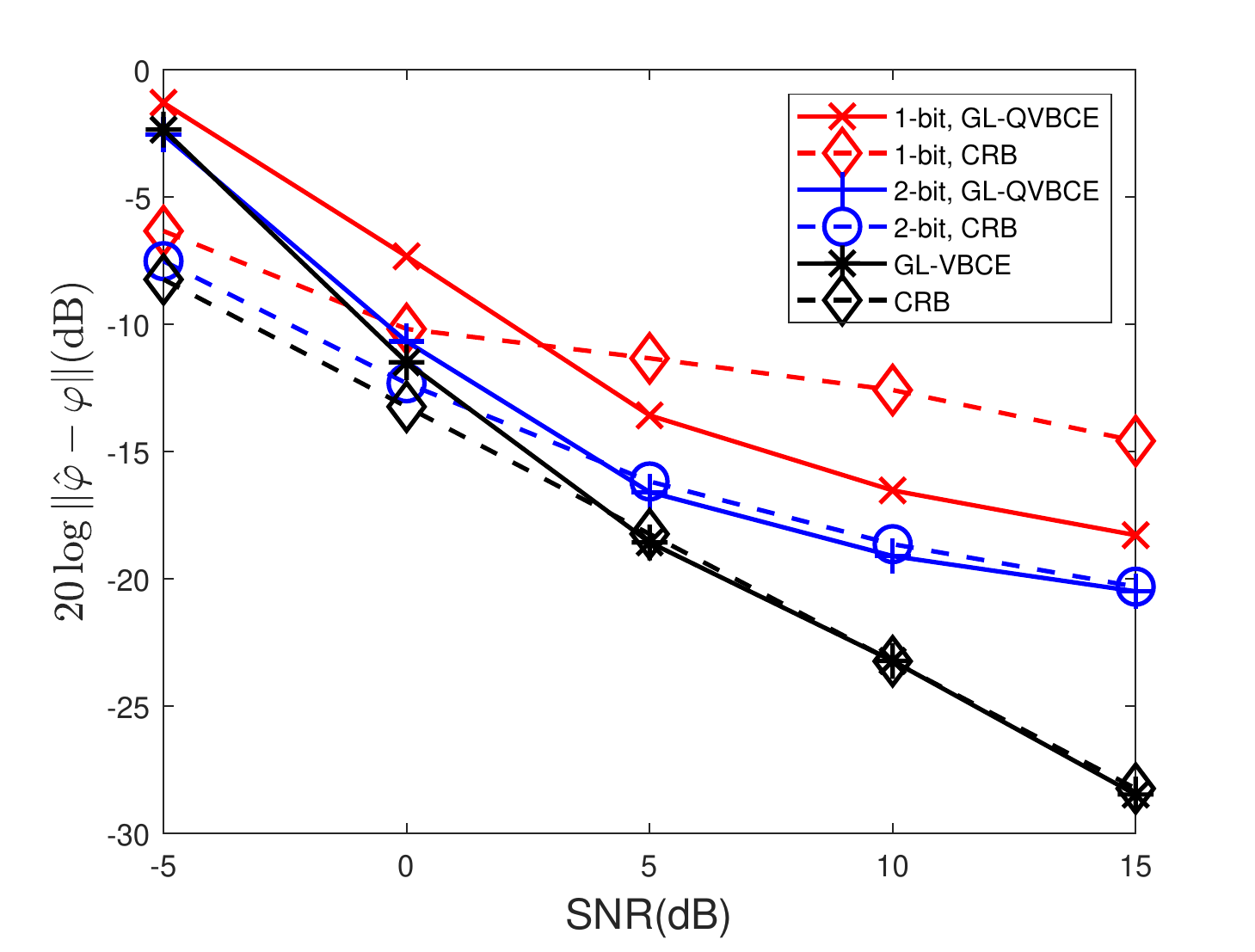}}

  \subfigure[]{
    \label{MSEtheta} %% label for second subfigure
    \includegraphics[width=2.8in]{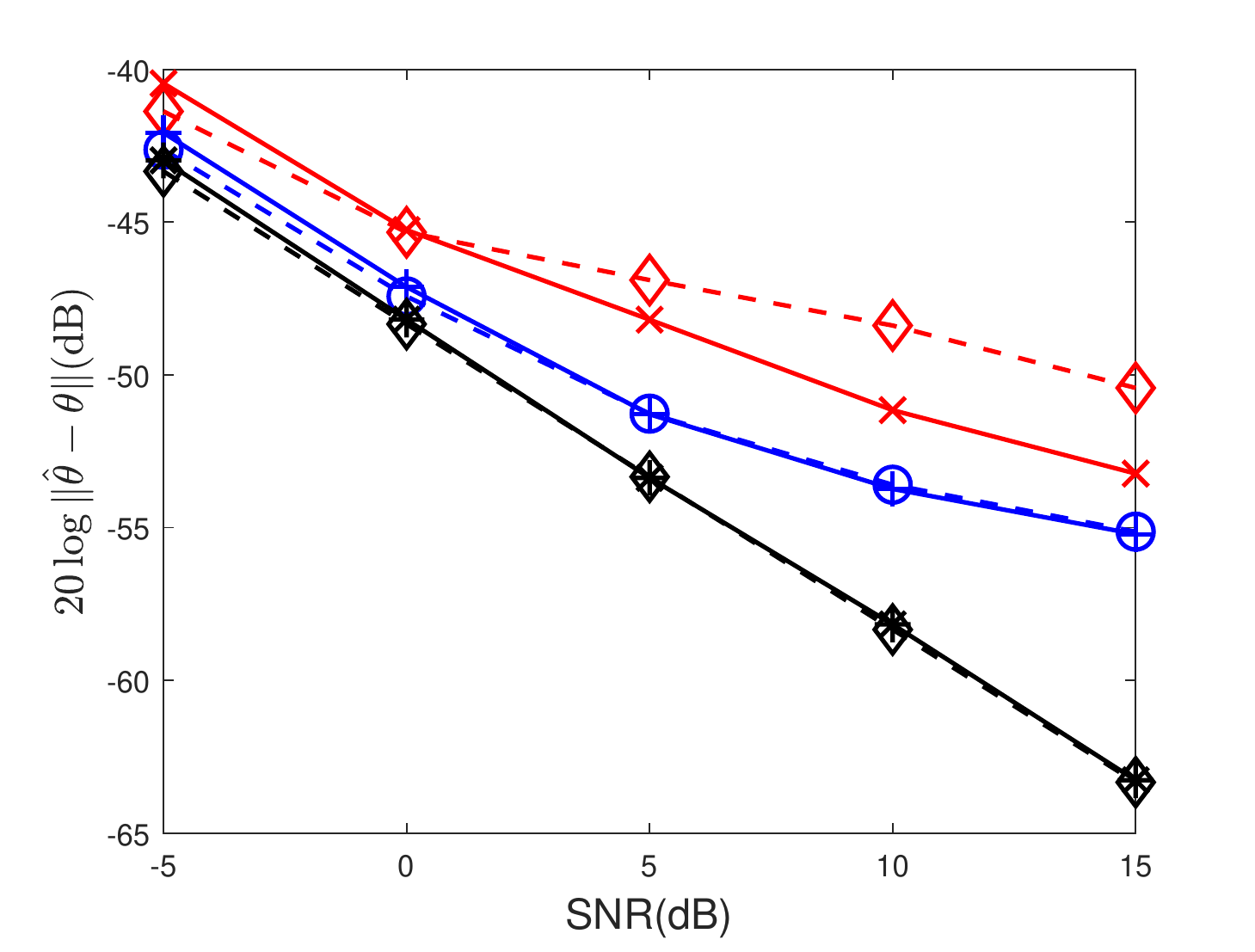}}
  \subfigure[]{
    \label{Prob} %% label for second subfigure
    \includegraphics[width=2.8in]{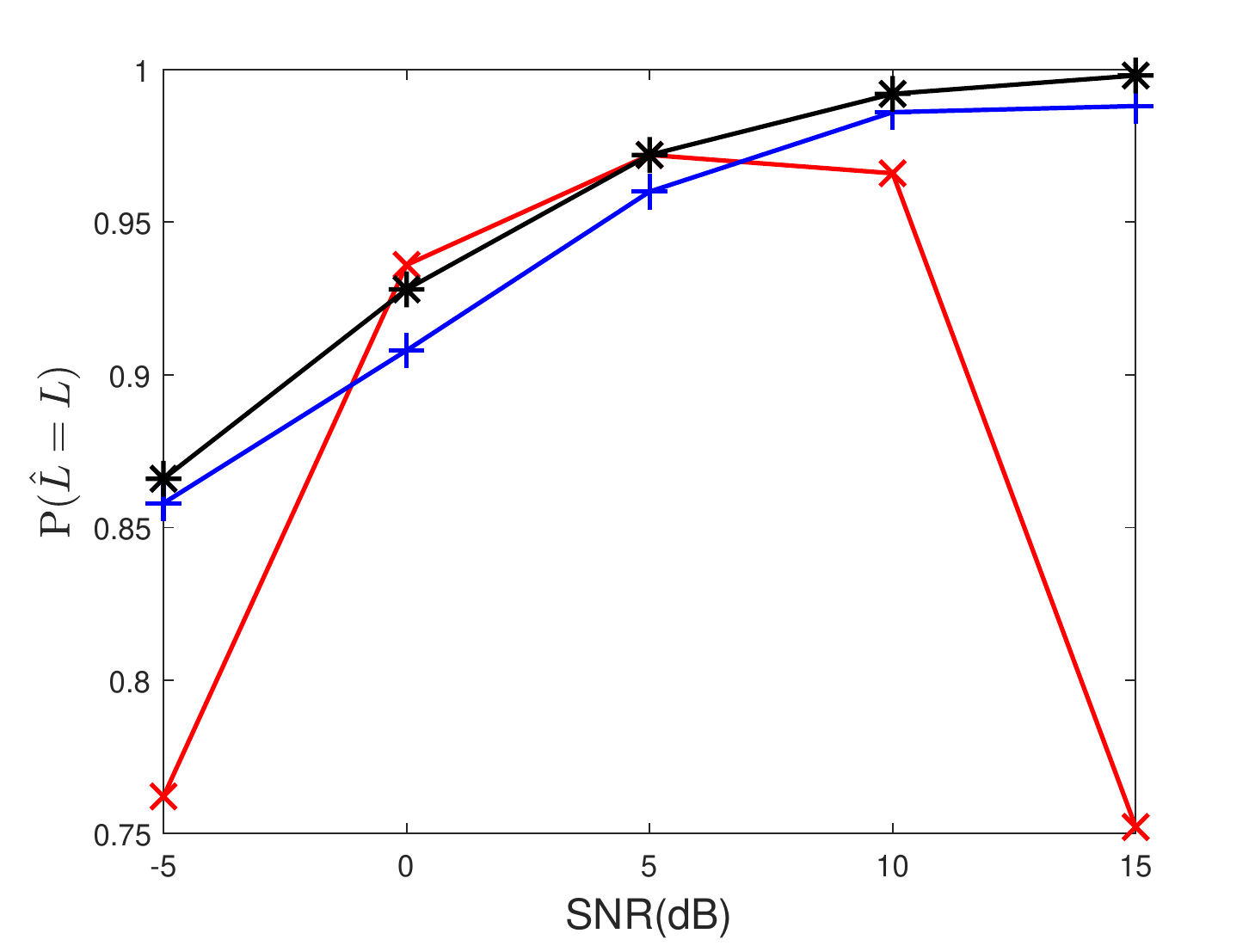}}
  \caption{(a) MSE of ray-gain estimation, (b) ray-phase estimation, (c) DOA estimation and CRBs versus SNRs for various channel estimators, probability of correct model order estimation. Results are averaged over $500$ MC trials.}
  \label{MSECRB} %% label for entire figure
\end{figure*}

The MSEs of ray-gain estimation, ray-phase estimation, DOA estimation against the CRB and the model order probability are presented in Fig. \ref{MSECRB}. It can be seen that all the algorithms except under one-bit quantization improve as SNR increases. On the one hand, noise is beneficial for recovering the magnitude information as magnitude information is lost under noiseless one-bit quantization. On the other hand, noise deteriorates the estimation performance as noise is unknown and adds uncertainty on the estimation. Thus it is expected that when SNR is low, the NMSE of the gain decreases with SNR, and when SNR is above a threshold, the MSE of the gain increases with SNR, as shown in Fig. \ref{MSEg} under one-bit quantization. For the phase and frequency estimation, the MSEs approach their CRBs under two bit quantization and unquantized setting. For the model order estimation probability, it increases as SNR increases for two bit quantization and unquantized setting. For one-bit quantization, when SNR exceeds $5$ dB, it decreases. The results demonstrate that for one-bit quantization, GL-QVBCE is more preferable under low SNR scenario.

\subsection{Effects of Bit-depth}
\begin{figure}
\centering{\includegraphics[width=2.8in]{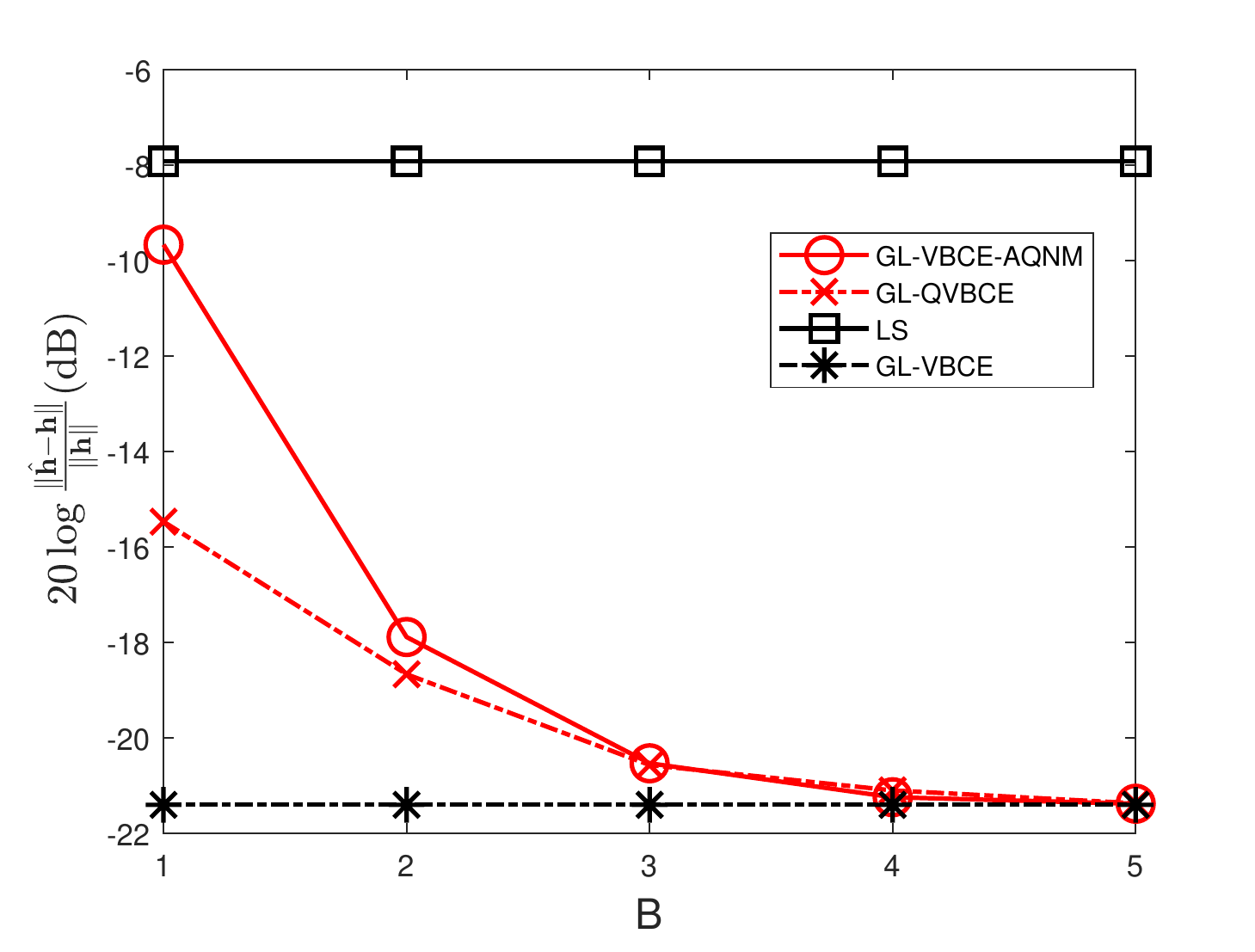}}
\caption{The mean NMSE of the channel versus the bit-depth for various algorithms. Here $N=M=64$, ${\rm SNR}=5$ dB, $T=2$.
}\label{MSE_BD}
\end{figure}

This experiment is conducted to illustrate the effect of bit-depth on the channel estimation and results is shown in Fig. \ref{MSE_BD}. It can be seen that increasing the bit-depth $B$ improves the channel estimation performance. As bit-depth increases, the performance gains decreases. From the figure, under 2-bit and 3-bit quantization, the performance gap between GL-QVBCE and GL-VBCE is about $3$ dB and $1$ dB. When $B\geq 4$, both GL-QVBCE and GL-VBCE-AQNM achieve almost identical performance with GL-VBCE under quantized setting.
\subsection{Effects of Number of Epffective Antennas}
\begin{figure}
\centering{\includegraphics[width=2.8in]{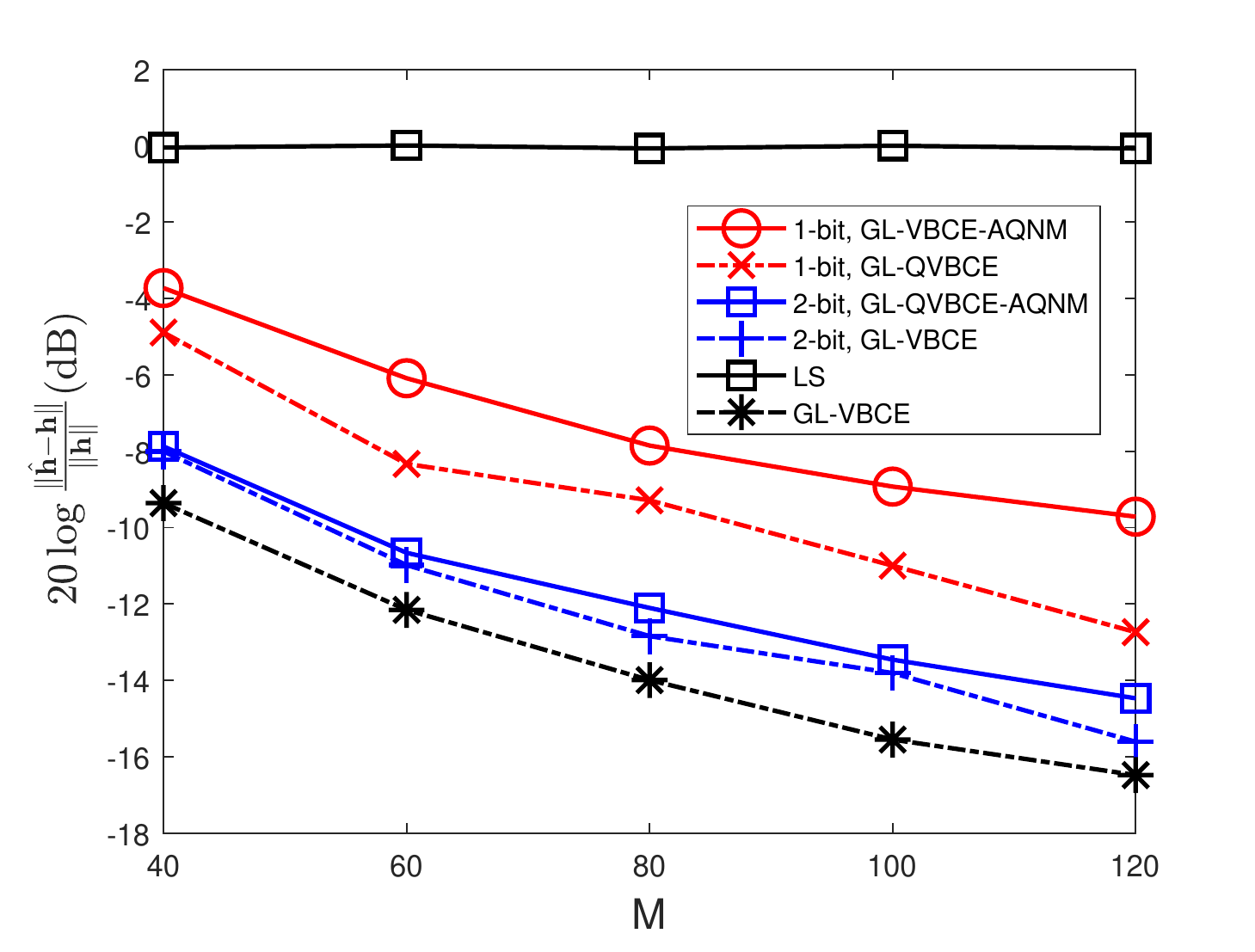}}
\caption{The mean NMSE of the channel versus the number of effective antennas for various algorithms. Here $N=200$, ${\rm SNR}=0$ dB, $T=1$.
}\label{MSE_M}
\end{figure}

Increasing the number of effective antennas will improve the performance of all the algorithms except LS. Fig. \ref{MSE_M} shows the NMSE of the channel versus the number of effective antennas $M$. It can be seen that under ${\rm SNR}=0$ dB, the NMSE of the LS approach is still $0$ dB as $M$ increases, which is consistent with the results (\ref{LSperb}). While for the GL-QVBCE and GL-VBCE approach, their performances improves with increasing $M$. Under 1-bit quantization, GL-QVBCE performs better than that of the AQNM model. While for 2-bit quantization, their performances are close.
\subsection{Effects of Multipath Numbers}
\begin{figure}
\centering{\includegraphics[width=2.8in]{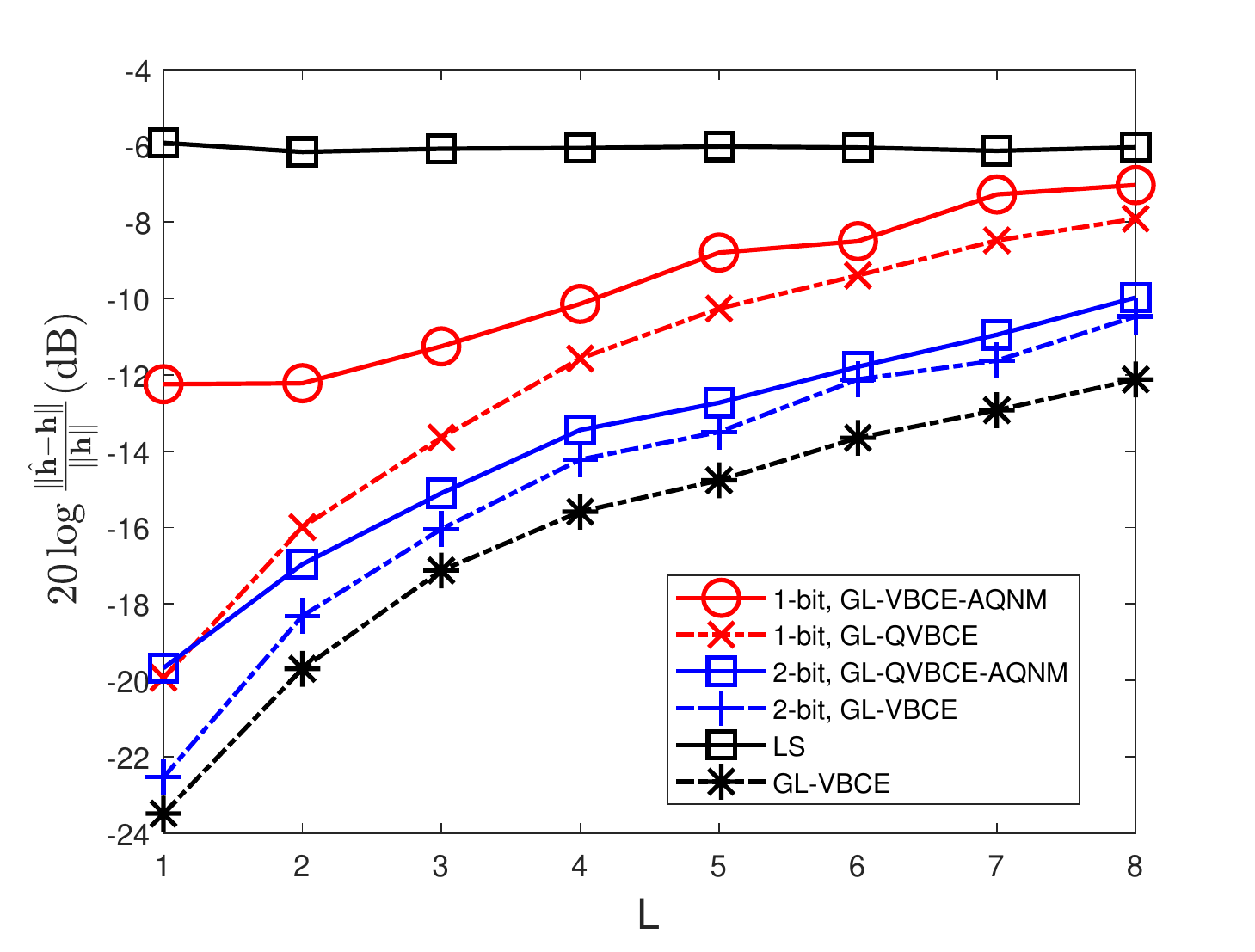}}
\caption{The mean NMSE of the channel versus the number of multipath for various algorithms. Here $N=M=64$, $T=4$, ${\rm SNR}=0$ dB.
}\label{MSE_L}
\end{figure}

The proposed approach utilizes the channel structure and thus improve the performance over traditional channel estimation algorithms. Fig. \ref{MSE_L} presents the MSE versus the number of multipath. It can be seen that performance of LS approach is stable irrespective of multipath numbers $L$, as (\ref{LSperb}) reveals. For all the algorithms exploiting the channel structure, their performance deteriorates as the number of multipath increases.
\subsection{Pilot Length}
\begin{figure}
\centering{\includegraphics[width=2.8in]{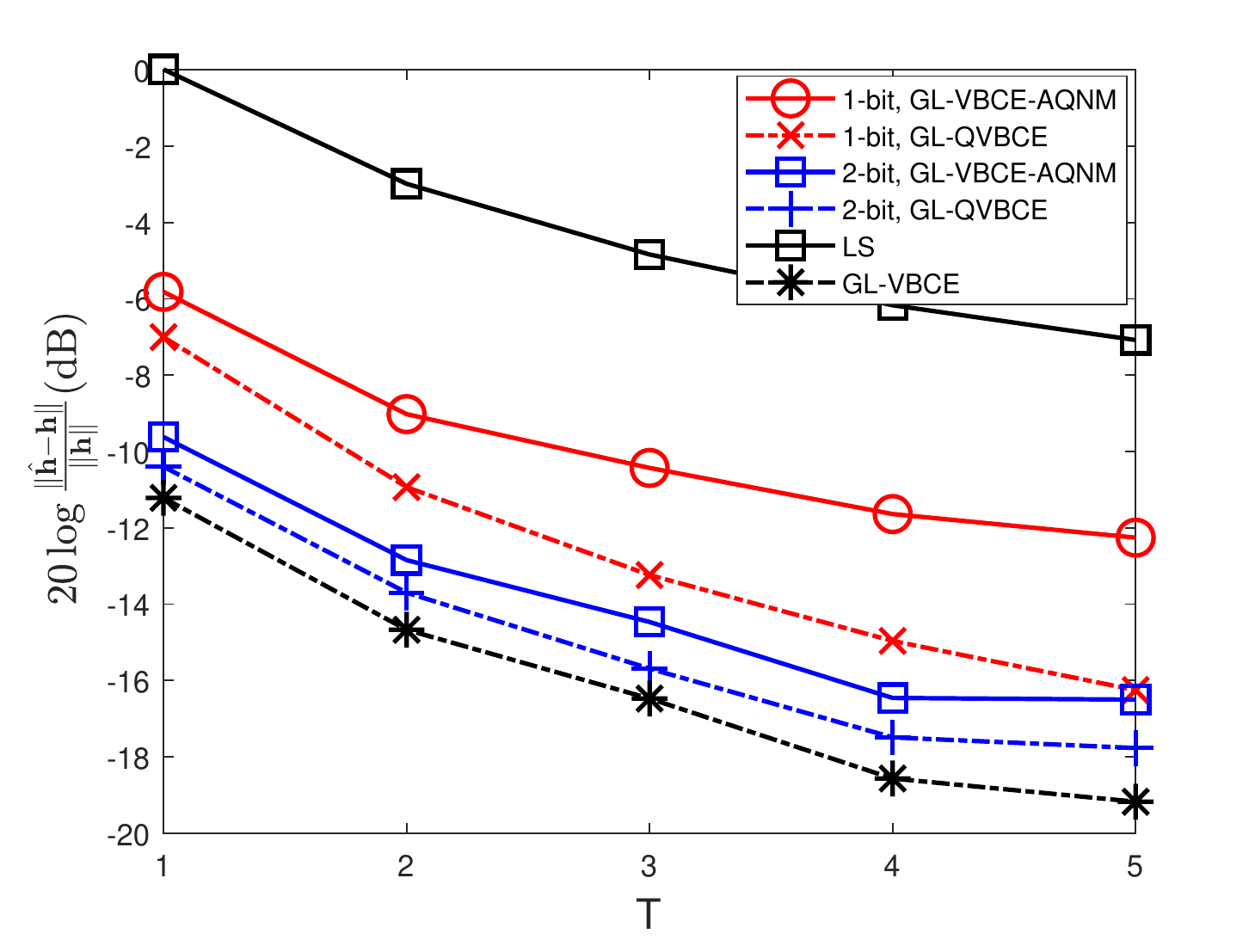}}
\caption{The mean NMSE of the channel versus the number of pilots for various algorithms. Here $N=M=64$, ${\rm SNR}=0$ dB.
}\label{MSE_T}
\end{figure}

An obvious fact is that the performance of the channel estimation algorithms should be improved by increasing the pilot length and Fig. \ref{MSE_T} validates this. In addition, it also shows similar phenomenon presented in subsection \ref{CEvsiter}.
\subsection{Sequential channel estimation}
\begin{figure}
\centering{\includegraphics[width=2.8in]{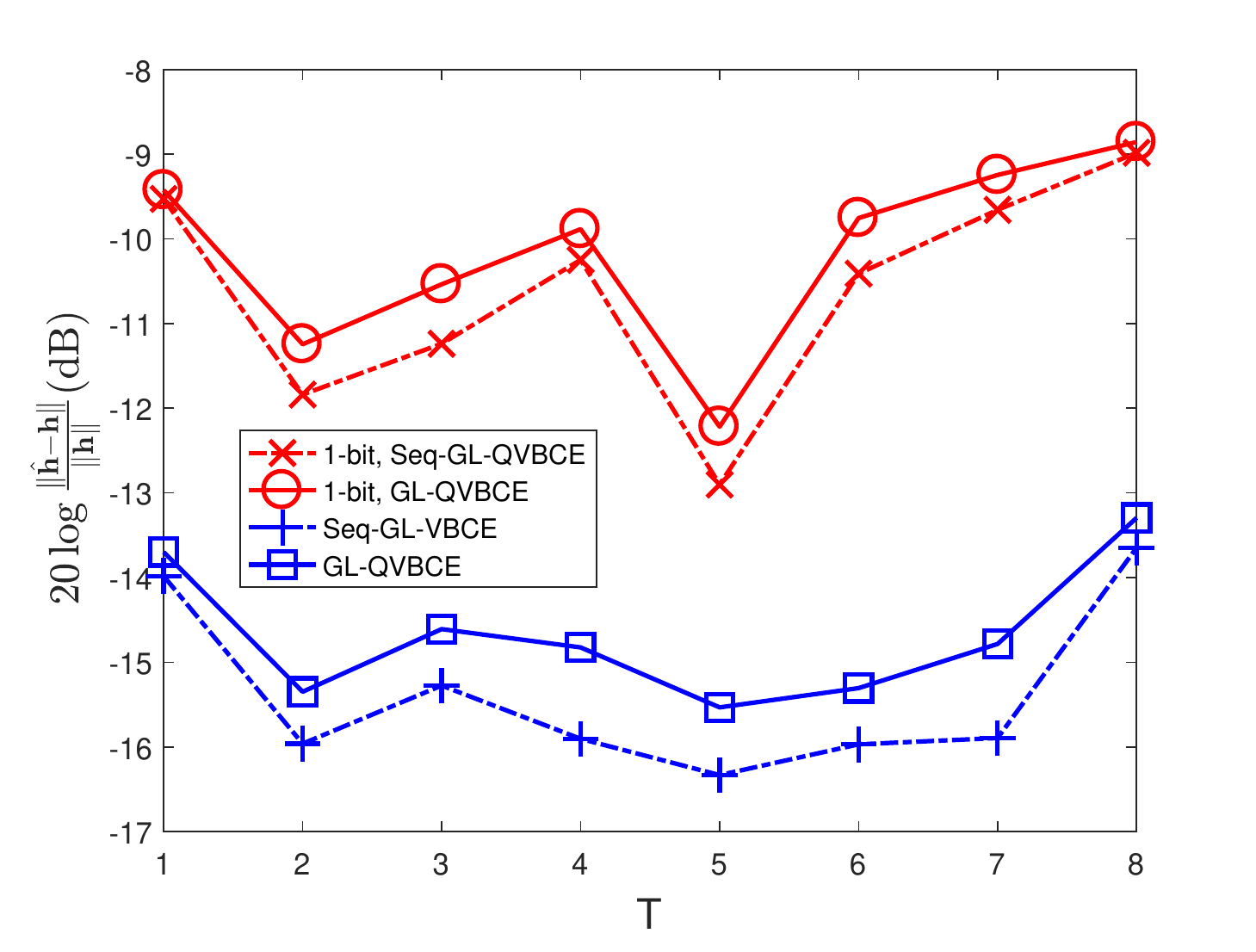}}
\caption{The mean NMSE of the channel versus the number of pilots for sequential estimation. Here $N=M=96$, $L=2$, ${\rm SNR}=0$ dB.
}\label{Seq_T}
\end{figure}

Consider a scenario where the DOAs are fixed across the pilots, while their complex coefficients are  completely independent. In this setting, Seq-GL-QVBCE and Seq-GL-VBCE algorithm are implemented to perform sequential channel estimation. As for the GL-QVBCE and GL-VBCE, we calculate the NMSE of channel for each pilot, separatively. We set $\lambda=0.1$ (\ref{lambdaset}) to damp the concentration parameter of the prior distribution. Results are presented in Fig. \ref{Seq_T}. It can be seen that both Seq-GL-QVBCE and Seq-GL-VBCE perform better than those of GL-QVBCE and GL-VBCE, respectively.
\section{Conclusion}
This paper proposes the GL-QVBCE and GL-VBCE algorithm exploiting channel structures to estimate the channel under quantized and unquantized setting. GL-QVBCE treats the DOAs as random parameters and performs the variational Bayesian estimation. When the DOAs are fixed across the pilots, Seq-GL-QVBCE and Seq-GL-VBCE are developed to perform sequential estimation. Numerical results are conducted to illustrate the effectiveness of the proposed algorithm.
%\newpage\newpage
\bibliographystyle{IEEEbib}
\bibliography{strings,refs}

\end{document}